\documentclass[sigconf]{acmart}
\newcommand\w{1} 

\AtBeginDocument{%
  \providecommand\BibTeX{{%
    \normalfont B\kern-0.5em{\scshape i\kern-0.25em b}\kern-0.8em\TeX}}}

\copyrightyear{2022}
\acmYear{2022}
\setcopyright{acmlicensed}\acmConference[CHI '22]{CHI Conference on Human Factors in Computing Systems}{April 29-May 5, 2022}{New Orleans, LA, USA}
\acmBooktitle{CHI Conference on Human Factors in Computing Systems (CHI '22), April 29-May 5, 2022, New Orleans, LA, USA}
\acmPrice{15.00}
\acmDOI{10.1145/3491102.3517647}
\acmISBN{978-1-4503-9157-3/22/04}

\usepackage{colortbl}
\usepackage{enumitem}

\def\markup{0}
\if\markup1
\newcommand{\rv}[1]{{\leavevmode\color{blue}#1}}
\else
\newcommand{\rv}[1]{#1}
\fi

\def\minor{1}
\if\minor1
\newcommand{\mc}[1]{{\leavevmode\color{blue}#1}}
\else
\newcommand{\mc}[1]{#1}
\fi

\begin{document}



\title[International Survey Study of Collaborative Data Analysis Practices]{\textit{``Merging Results Is No Easy Task''}: An International Survey Study of Collaborative Data Analysis Practices Among UX Practitioners}

\author{Emily Kuang}
\affiliation{%
  \institution{School of Information}
  \institution{Rochester Institute of Technology}
  \city{Rochester}
  \state{New York}
  \country{USA}
}
\email{ek8093@rit.edu}

\author{Xiaofu Jin}
\affiliation{%
  \institution{Computational Media and Arts}
  \institution{The Hong Kong University of Science and Technology (Guangzhou)}
  \city{Guangzhou}
  \country{China}
}
\email{xjinao@connect.ust.hk}

\author{Mingming Fan}
\authornote{Corresponding Author}
\affiliation{
  \institution{Computational Media and Arts}
  \institution{The Hong Kong University of Science and Technology (Guangzhou)}
  \city{Guangzhou}
  \country{China}
}
\affiliation{
  \institution{Integrative Systems and Design \& Computer Science and Engineering}
  \institution{The Hong Kong University of Science and Technology}
  \city{Hong Kong SAR}
  \country{China}
}
\email{mingmingfan@ust.hk}






\renewcommand{\shortauthors}{Kuang, Jin, and Fan}

\begin{abstract}

Analysis is a key part of usability testing where UX practitioners seek to identify usability problems and generate redesign suggestions. Although previous research reported how analysis was conducted, the findings were typically focused on \textit{individual analysis} or based on a small number of professionals in specific geographic regions.
We conducted an online international survey of 279 UX practitioners on their practices and challenges while \textit{collaborating during data analysis}. 
We found that UX practitioners were often under time pressure to conduct analysis and adopted three modes of collaboration: independently analyze different portions of the data and then collaborate, collaboratively analyze the session with little or no independent analysis, and independently analyze the same set of data and then collaborate.
Moreover, most encountered challenges related to lack of resources, disagreements with colleagues regarding usability problems, and difficulty merging analysis from multiple practitioners. We discuss design implications to better support collaborative data analysis. 
\end{abstract}

\ccsdesc[500]{Human-centered computing~Empirical studies in collaborative and social computing}

\keywords{User experience; UX; Usability testing; Data analysis; Collaboration; Survey}


\maketitle

\section{Introduction}

Usability testing is a frequently employed user-centred design method for detecting usability problems \cite{vredenburg_survey_2002,fan_practices_2020}.
Despite being effective and useful, analyzing usability test sessions can be tedious, challenging, and time-consuming~\cite{norgaard_what_2006, chilana_understanding_2010, folstad_analysis_2012,fan_practices_2020}. 
When analyzing usability test sessions, user experience (UX) practitioners often need to attend to multiple behavioral signals in both the visual and audio channels of the recordings and conduct several tasks (e.g., look at user's actions, listen to user's verbalizations, and write their own annotations) simultaneously~\cite{chilana_understanding_2010}.
Furthermore, UX practitioners working in industry often face time pressure to deliver their analysis results~\cite{folstad_analysis_2012,mcdonald_exploring_2012,fan_practices_2020}. Consequently, it is not uncommon that UX practitioners might miss important usability problems or misinterpret them \rv{\cite{hertzum_evaluator_2001, fan_vista_2020}}.

To address these challenges, UX practitioners are recommended to analyze data \textit{collaboratively} to increase the completeness and reliability~\cite{folstad_analysis_2010, folstad_analysis_2012,fan_practices_2020}. By bringing different perspectives from multiple practitioners together, collaboration could help overcome the \emph{``evaluator effect''}~\cite{hertzum_evaluator_2001}---a common phenomenon where different UX practitioners may uncover or interpret usability problems and their severity levels differently. Unfortunately, fewer than 30\% of UX practitioners have a chance to collaborate with others to analyze the same usability test session due to practical constraints, such as limited resources~\cite{folstad_analysis_2012}. Even if they do collaborate, they often have to put together an ad-hoc set of tools to support their analysis of test sessions and collaboration~\cite{fan_practices_2020,mcdonald_exploring_2012}.

To better understand how UX practitioners work in practice, researchers conducted various studies in the past decades~\cite{folstad_analysis_2010,norgaard_what_2006,shi_field_2008,mcdonald_exploring_2012,fan_practices_2020}. 
While informative, these studies were typically based on small data samples (e.g., interview with eleven UX practitioners \cite{folstad_analysis_2010}) or focused on the usage of specific methods, such as think-aloud protocols~\cite{mcdonald_exploring_2012,fan_practices_2020}.
For the few studies that did focus on general practices of UX practitioners (e.g., \cite{folstad_analysis_2012}), they offered limited insights into collaboration practices and purposes around data analysis among UX practitioners. Furthermore, almost all previous studies were conducted a decade ago except Fan et al.'s study~\cite{fan_practices_2020}, which was focused on think-aloud protocol usage instead of general practices. In the most recent decade, new technology breakthroughs, such as deep learning and virtual reality (VR), have spurred a variety of AI-enabled products (e.g., smart speakers) and VR applications. These emerging products enable new interaction methods and provide novel user experiences, which might require UX practitioners to adapt their analysis and collaboration practices. Indeed, new UX analytical platforms (e.g., UserTesting.com~\cite{usertesting_usertesting_2021} and FullStory~\cite{fullstory_fullstory_2021}) and data-informed design and analytical tools (e.g., MixPanel~\cite{mixpanel_2021}), have also emerged in the meantime. As a result, there is a lack of understanding of how UX practitioners collaborate when analyzing usability test sessions in today's social and technological contexts.
In addition, it is important to gain an understanding of the practices used during independent analysis as it lays the foundation for collaboration \cite{folstad_analysis_2010}.
We take a first step to address this gap by exploring the following research questions (RQs):

\begin{itemize}
\item {RQ1}: What are the independent data analysis practices of UX practitioners? 
\item {RQ2}: What are the practices, challenges, and desired improvements for collaborative data analysis of UX practitioners? 
\item {RQ3}: How might UX practitioners' data analysis and collaboration practices be affected by their years of experience and team size? 
\end{itemize}

To answer these RQs, we conducted an online survey study with a representative sample of 279 UX professionals who had different levels of UX experience from six continents in the world. 
The survey focused on the resources they used when analyzing usability test recordings and how they collaborated with fellow UX team members. 
The survey also collected new features that they hoped to have in new collaborative tools to better support their collaboration in the data analysis phase. 

Our results show that about two thirds (66\%) of UX practitioners were under time pressure to conduct analysis and most of them utilized structured formats and severity ratings when describing usability problems. 
The top three purposes of collaboration were to identify more usability problems, generate redesign suggestions, and improve reliability respectively. We also found that respondents with greater years of UX experience were more likely to consider the improvement of reliability when collaborating.
Their collaboration happened in a variety of circumstances and most of them had encountered challenges related to lack of time, disagreements with colleagues regarding usability problems, and difficulty merging analysis from multiple practitioners. 
\rv{
Based on the challenges that respondents encountered, we derive some design considerations that may benefit both UX practitioners and researchers by suggesting target areas for collaboration improvement and pointing out avenues for future research. 
}

In sum, we make the following contributions in this work:
\begin{itemize}
\item We present a quantitative understanding of UX practitioners' collaboration practices and challenges based on a sample of 279 international UX professionals; 
\item We show potential correlations between their years of UX experience and team size with their analysis practices; 
\item We highlight design considerations for improving data analysis and collaboration among UX practitioners. 
\end{itemize}

\section{Background and Related Work}

Our work is motivated and informed by prior work on \textit{UX analysis practices}, and \textit{collaborative data analysis among UX practitioners}.
\subsection{Analysis of Data Collected from Usability Test Sessions}
Usability is defined as the extent to which a system enables users to achieve goals effectively, efficiently and with satisfaction, based on the intended context of use \cite{iso_iso_2020}. 
To improve system usability, UX practitioners must identify and address usability problems that users may encounter. Examples of usability evaluations include observations, cognitive walkthroughs, interviews, focus groups, heuristic evaluations, and usability testing \cite{nielsen_usability_1994}. 
The most frequently employed method for detecting such problems is through usability testing \cite{fan_practices_2020}. This is due to strong evidence that usability testing is effective in a wide variety of usage scenarios \cite{norgaard_what_2006}.

While usability testing has been shown to be effective and useful across various methods and domains, conducting sound usability testing is challenging. 
In addition to the challenges practitioners face during data collection \cite{dumas_moderating_2008, norgaard_what_2006}, they need to use a variety of tools to record the session for later analysis.
All participants in an exploratory study on usability testing reported they video-record the session and take notes during the session \cite{folstad_analysis_2010}. 
Similarly, in other surveys of UX practitioners, 100\% \cite{folstad_analysis_2012} and 89\% \cite{fan_practices_2020} of the participants take observations notes, 73\% \cite{folstad_analysis_2012} and 77\% \cite{fan_practices_2020} of the participants video-record the session, and 70\% \cite{fan_practices_2020} conduct post-task interviews. 


During analysis, UX practitioners use these diverse sources of data to generate coherent descriptions of usability problems, which include possible causes, effects, and solutions \cite{folstad_analysis_2010}, following both light-weight and formal analysis procedures.
The Instant Data Analysis (IDA) technique is a light weight approach requiring practitioners to perform quick analysis immediately upon the conclusion of a usability testing session \cite{kjeldskov_instant_2004}. 
On the other hand, the User Action Framework (UAF) \cite{andre_user_2001} and the Structured Usability Problems EXtraction (SUPEX) framework \cite{cockton_framework_1999} describe formal standards for analysis.
What's more, templates \cite{cockton_reconditioned_2004, lavery_comparison_1997} and guidelines \cite{capra_usability_2006} have been developed for organizing problem descriptions, prioritizing the problems (also known as the severity rating \cite{lewis_usability_2006}), and deriving potential interventions. 
However, in practice, the analysis is often unstructured, incomplete with no identification of causes or solutions \cite{norgaard_what_2006}, and is informal and lack rigor~\cite{folstad_analysis_2010}. 
Even if practitioners use structured formats, they tend to use homegrown formats developed by themselves or their companies~\cite{folstad_analysis_2012}.

In sum, there seems to be a gap between recommended guidelines and practices for UX data analysis. This might induce vague evaluation procedures that lead to bias or unclear problem criteria, which result in an overflow of reported issues \cite{hertzum_evaluator_2001}. As new methods and tools have emerged since the majority of the studies regarding analysis practices were conducted a decade ago (e.g., \cite{norgaard_what_2006,folstad_analysis_2010,folstad_analysis_2012,mcdonald_exploring_2012}), we seek to understand the current practices adopted by UX practitioners when performing independent data analysis, which is usually done prior to collaborating with other practitioners~\cite{folstad_analysis_2010}. 

\subsection{Collaborative Data Analysis among UX Practitioners}
While collaboration during usability testing happens under various circumstances, such as between a group of participants \cite{de_vreede_repeatable_2005}, between UX practitioners and stakeholders \cite{simpson_practice_1991, fruhling_collaborative_2006}, and between UX practitioners and experts of complex domains \cite{chilana_understanding_2010}, this paper addresses collaboration in the context of UX practitioners working with their colleagues during the data analysis stage.

\textbf{The Need for Collaboration.} Analyzing usability test recordings is challenging and time-consuming \cite{norgaard_what_2006, chilana_understanding_2010, folstad_analysis_2012, fan_practices_2020}. 
UX practitioners need to pay attention to both the visual and audio channels of the recording while observing user's actions and writing notes simultaneously \cite{chilana_understanding_2010}. 
However, in practice, they only have limited time and resources to complete analysis \cite{folstad_analysis_2012}. 
As a result, they have to juggle between achieving high reliability and validity in their analysis and completing their analysis efficiently \cite{fan_practices_2020}. 
One important way to cope with the tension is to collaborate with other UX practitioners to review test sessions together \cite{fan_practices_2020}, which not only divides the workload but is also perceived to improve the thoroughness of problem detection \cite{sears_heuristic_1997}.

Another benefit of collaboration is alleviating a persistent issue encountered by UX practitioners in UX data analysis---the \textit{``evaluator effect,''} which refers to the fact that different practitioners tend to identify different sets of UX problems even when analyzing the same test session \cite{hertzum_evaluator_2001,jacobsen_evaluator_1998}.
This issue exists for novice and experienced practitioners, for problem detection and severity assessment, and for simple and complex systems \cite{hertzum_evaluator_2001}. 
The majority of practitioners indeed perceived the \textit{``evaluator effect''} when merging their individual findings into group evaluations \cite{hertzum_what_2013}. 
Collaboration among UX practitioners enables them to view the data from different perspectives, which could lead to an increase in both the reliability \cite{hertzum_evaluator_2001} and thoroughness of the problems identified \cite{sears_heuristic_1997}. 
A study has shown that practitioners working in isolation identified significantly fewer usability problems than a team of three to five evaluators \cite{nielsen_mathematical_1993}.
In other words, collaboration allows a higher possibility that all the usability problems in a session can be found and that the analysis is more consistent across practitioners. 

\textbf{Collaboration Practices}. 
Although many UX practitioners recognized the value of collaboration in improving reliability, few collaborated with others when analyzing the same test recording in practice \cite{folstad_analysis_2012}. 
In an exploratory study with eleven UX practitioners, only three reported having some form of collaboration \cite{folstad_analysis_2010}. 
More recently, a survey of 197 UX practitioners found that more than half (56\%) of the participants analyzed data and wrote informal reports \textit{alone} and 42\% also analyzed data and wrote formal reports \textit{alone} \cite{fan_practices_2020}. 
One important reason for limited collaboration is that UX practitioners often work on projects that have quick turnaround cycles or limited resources~\cite{fan_practices_2020,mcdonald_exploring_2012,chilana_understanding_2010}. 

Another key part of collaboration between UX practitioners is the use of software tools to facilitate session review and analysis. 
Collaborative software has been shown to support heuristic evaluation in groups, where it allowed groups of evaluators to reduce the number of duplicate usability problems, reach consensus earlier, and improve productivity \cite{lowry_improving_2003}.
Usability testing in general also benefits from collaborative software support. However, this software mainly served as a collaborative writing tool and did not provide any session review functionalities \cite{lowry_users_2002}. Moreover, it required the use of an external meeting software for distributed collaboration, which demonstrates the lack of \textit{integrated} tools that support both analysis and communication among UX practitioners. 

Numerous commercial tools are also used by UX practitioners in industry. 
Although \textit{offline tools}, such as Morae \cite{techsmith_morae_2020}, Noldus Viso \cite{noldus_record_2020}, and Silverback \cite{silverback_silverback_2019}, can be installed on a local machine and allow for reviewing sessions with functionalities like note-taking and marking events on video progress bar, they do not support collaboration well. In contrast, \textit{online tools} (e.g. UserTesting~\cite{usertesting_usertesting_2021} and FullStory~\cite{fullstory_fullstory_2021}) allow for more flexible collaboration, where multiple UX practitioners can gain access to the same session recordings. 
However, such tools provide limited data analysis capabilities, which are comprised of session playback, note-taking, and tagging, and practitioners still need to communicate through an external platform to discuss their analysis results. 
Consequently, UX practitioners often have to put together a generic set of tools on an ad hoc basis that are not designed specifically for their needs \cite{fan_practices_2020, mcdonald_exploring_2012}, such as screen recorders, eye-trackers, prototyping tools, web analytics tools, spreadsheets, text editors, and presentation tools \cite{folstad_analysis_2012}, to collaboratively capture and analyze various sources of data. 

In sum, despite the necessity of collaboration, UX practitioners face many challenges for collaboration in practice. Although previous research reported how UX practitioners conducted analysis on user evaluations in general \cite{fan_practices_2020, folstad_analysis_2010, folstad_analysis_2012, boren_thinking_2000}, the findings did not provide specific insights into practices and factors affecting collaboration and its associated challenges. Therefore, to help identify opportunities to improve collaboration among UX practitioners, we conducted a survey study to first understand UX practitioners' practices during independent analysis (RQ1), then their practices, challenges, and desired improvements for collaborative analysis in detail (RQ2), and finally how their years of experience and team size might affect analysis and collaboration practices (RQ3). 
\section{Method}

We employed a web-based questionnaire survey to reach a broader set of UX practitioners globally since analysis practices are expected to vary among practitioners. This method allowed us to get a diverse sample of respondents at different stages of their career across six continents. Our choice of method is in line with previous survey studies of usability practices \cite{folstad_analysis_2010, mcdonald_exploring_2012,  fan_practices_2020}. Below we present details about our survey design, recruitment methods, and respondents. 

\begin{table*}
  \caption{Summary of respondent demographics (N = 279)}
  \label{tab:demographics}
  \begin{tabular}{lr|lr}
    \hline
    \rowcolor[gray]{0.9}
    Gender & & Size of UX Team & \\
    \hline
    Female & 141 (50.5\%) & 1 person & 84 (30.1\%) \\
    Male & 115 (41.2\%) & 2 - 5 people & 148 (53.1\%) \\
    Prefer not to say & 22 (7.9\%) & 6 - 10 people & 26 (9.3\%) \\
    Prefer to self-describe & 1 (0.4\%) & 11 and above & 21 (7.5\%) \\
    \hline
    \rowcolor[gray]{0.9}
    Work Location & & Years of Experience in UX & \\
    \hline
    North America & 119 (42.7\%) & Less than 1 year & 77 (27.6\%) \\
    Asia & 70 (25.1\%) & 1-2 years & 74 (26.5\%) \\
    Europe & 64 (22.9\%) & 3-5 years & 69 (24.7\%) \\
    South America & 21 (7.5\%) & 6-9 years & 30 (10.8\%) \\
    Africa & 3 (1.1\%) & 10 or more years & 29 (10.4\%) \\
    Oceania & 2 (0.7\%) &  &  \\
    \hline
\end{tabular}
\end{table*}

\subsection{Survey Design}
To address the research questions, the survey was separated into three sections: (1) current independent data analysis practices, (2) practices and challenges of collaboration, and (3) new features for collaboration tools to assist with data analysis. 
The questions sought to understand background data (practitioner's work experience and evaluation context), practices and tools used for usability problem identification, and perceived challenges and desired improvements to support collaboration practices. 
The majority of survey questions were fixed response items, such as single choice, multiple-choice (i.e. select-all-that-apply), and Likert-scale ratings, which were required. 
To supplement the fixed responses and gain a deeper understanding, there were eight optional free text questions that sought explanations for why practitioners chose certain responses. These questions also asked about what are the greatest challenges practitioners encounter, how they handle disagreements with colleagues, and how their collaboration could be better supported. For the free response questions, the average length of the answers was three sentences. They were first coded independently by two authors before meeting to discuss and group them into themes. 
The questions in sections (2) and (3) regarding collaboration practices and challenges were not shown to respondents who answered "1 person (just myself)" for the size of UX team since those questions were not applicable to their work experiences. 

\subsection{Testing the Survey Tool}
To pretest the survey, We followed Dillman's suggested three stage process \cite{dillman_mail_2000}. 
First, the survey was reviewed by colleagues to uncover potential misinterpretations and additional questions that may have been overlooked. 
Next, we discussed the survey plan with UX practitioners to ensure the sufficient motivation for the survey as well as communication clarity. 
Finally, we performed pilot-testing with four UX practitioners to identify any flaws in the survey questions and distribution platform to ensure that the length is appropriate. 

\subsection{Respondent Recruitment}
The survey was approved by the authors' institutional review board (IRB) and was deployed in late 2020. It was conducted through Qualtrics and was accessed by all respondents through an anonymous link. The respondents were required to be UX practitioners with prior experience conducting usability evaluations. The respondents were recruited through several channels; invitations were distributed via mailing lists of local UXPA (i.e., User Experience Professional Association) chapters around the world (which requires qualification verifications for their members), UX Professional groups on LinkedIn, and industry connections. 
As incentives to participate in the survey, all respondents who recorded their e-mail address could get early access to the project results and were included in a lottery for \$30.

\subsection{Data Quality}
We took multiple precautions to ensure data quality. We provided plain descriptions of terms to explain the options, such as ``collaborative writing: team members can make annotations on same recording''. 
Respondents were informed that data collection was fully anonymous to avoid bias. Respondents who chose to leave their email address \mc{for} the lottery were assured that the e-mail addresses would be separated from the questionnaire data. 
\mc{We also} discarded certain responses from the total 309 attempts of the questionnaire to ensure data quality. 
We excluded 30 responses for leaving after the first few questions (26), declining to provide consent (3), and selecting "Strongly Agree" across all Likert scales, which suggests that each statement was not read carefully (1). 

\subsection{Respondent Demographics}


We received a total of 279 valid responses. 
Average survey completion time was 6.8 minutes ($Md\footnote{Md = median} = 6.2$). 
Table \ref{tab:demographics} shows a summary of respondents' demographics and work experience. 
They worked across six continents with the largest number of respondents in North America, which was likely because it shared the largest number of companies and was where the survey was distributed from. Nonetheless, there were reasonable numbers of respondents in Asia and Europe. One limitation was the small number of responses from other continents, which will be discussed in Sec.~\ref{sec:limitations}. 

We received a near equal distribution for respondents with less than 1 year, 1-2 years, and 3-5 years of prior work experience, making up a total of 79\% with 5 years or less of experience working in UX. 
Over half (53\%) of the respondents worked on teams consisting of 2-5 people, with only 17\% reporting they worked with teams greater than 5 people. 
Interestingly, 30\% of the respondents were the only UX practitioner on their team.

\section{Findings}

In the process of analyzing usability testing sessions, it is typical that individual analysis is done prior to collaboration \cite{folstad_analysis_2010}. 
Thus, it is important to gain an understanding of the practices used in the individual portion. 
In this section, we will first discuss their independent analysis practices (RQ1), then we will discuss in greater detail their collaboration (RQ2), and correlations with prior UX experience and team size (RQ3). 

\subsection{RQ1: Independent Data Analysis Practices}
Our analysis revealed two themes about individual data analysis practices: \textit{frequency and time spent on usability testing} and \textit{resources used for usability problem identification}.

\subsubsection{Frequency and Time Spent on Usability Testing}
\begin{figure}[h]
  \centering
  \includegraphics[width=\w\linewidth]{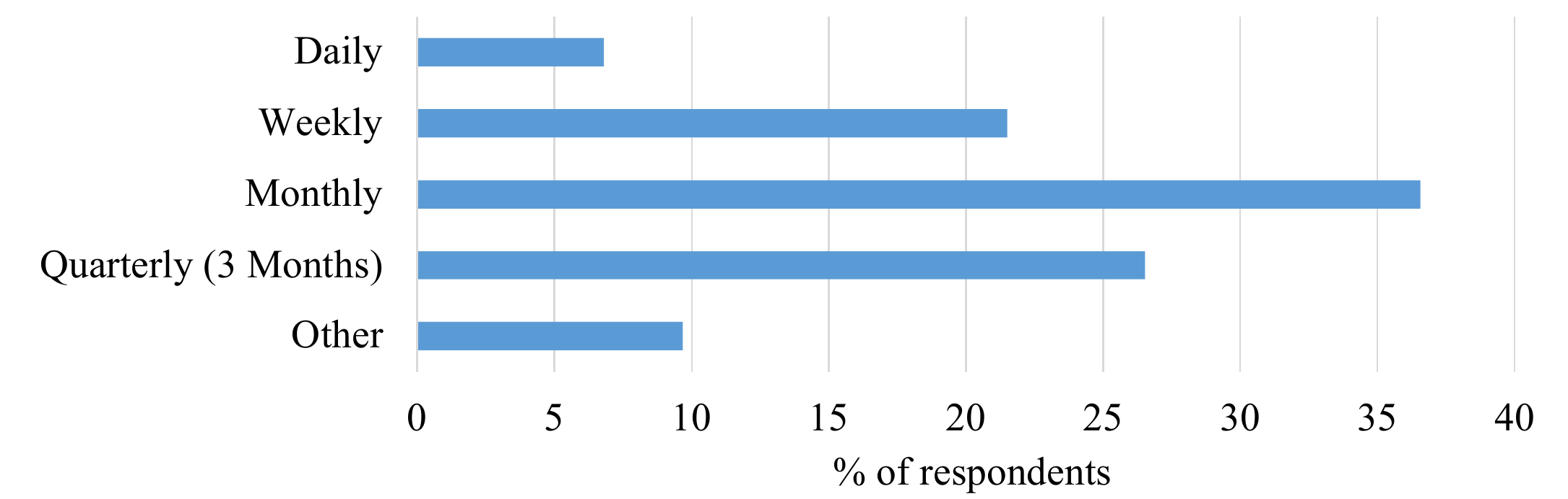}
  \caption{Frequency of conducting usability testing $(N=279, single-select)$}
  \Description{Bar graph of the frequency of conducting usability testing where the most common choice is monthly}
  \label{fig:Frequency}
\end{figure}

Figure \ref{fig:Frequency} shows that the highest portion of the respondents conducted usability testing monthly (37\%), followed by quarterly (27\%), and weekly (22\%). 
Among the 24 respondents who chose ``Other'', over half of them (59\%) reported that the frequency varies depending on the schedule of the projects and that some projects may need multiple rounds of testing.

\begin{figure}[h]
  \centering
  \includegraphics[width=\w\linewidth]{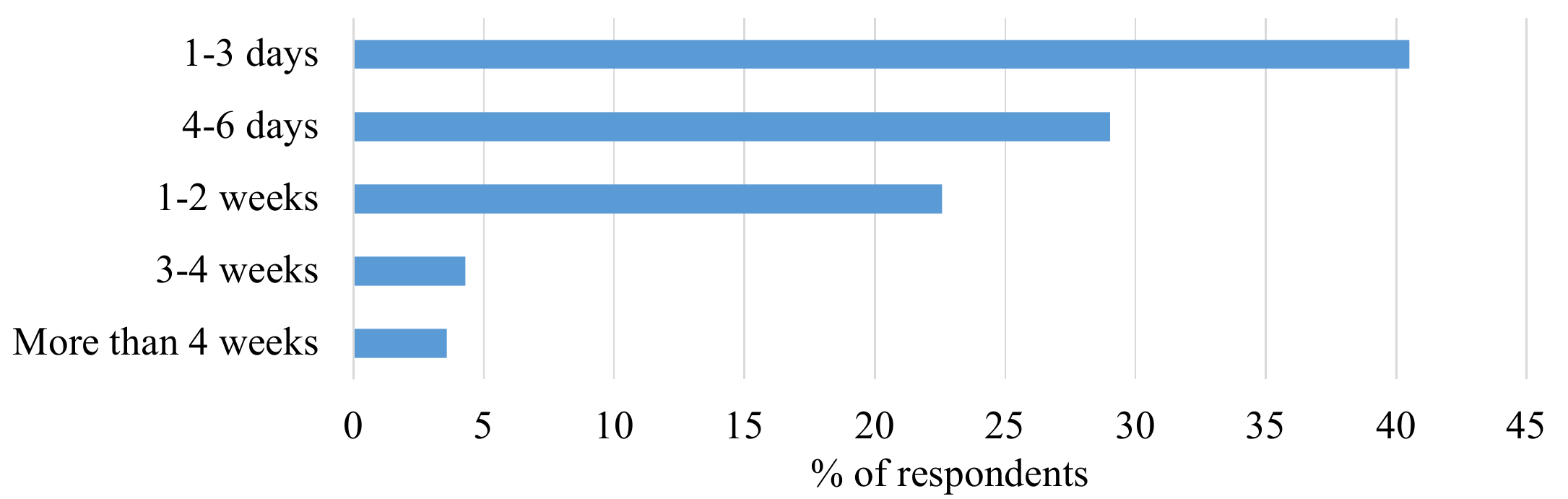}
  \caption{Time spent on analyzing data collected from usability testing sessions $(N=279, single-select)$}
  \Description{Bar graph of the time spent on data analysis where 70\% of respondents needed to complete it within one week}
  \label{fig:TimeCost}
\end{figure}

Figure \ref{fig:TimeCost} shows that 41\% of the respondents only had \textit{1-3 days} to analyze data collected from usability testing sessions and 29\% of them had \textit{4-6 days}. 
Since the majority (70\%) of the respondents needed to complete their analysis within one week of the usability testing session, there is significant time pressure on most UX practitioners. 
Indeed, a similar proportion (66\%) of respondents ``somewhat agreed'' or ``strongly agreed'' that they found it \textit{time-consuming to conduct data analysis} (Md=4, IQR\footnote{IQR = inter-quartile range, which is the difference between the 75th and 25th percentiles}=1). 

\subsubsection{Resources Used for Usability Problem Identification}
\begin{figure}[h]
  \centering
  \includegraphics[width=\w\linewidth]{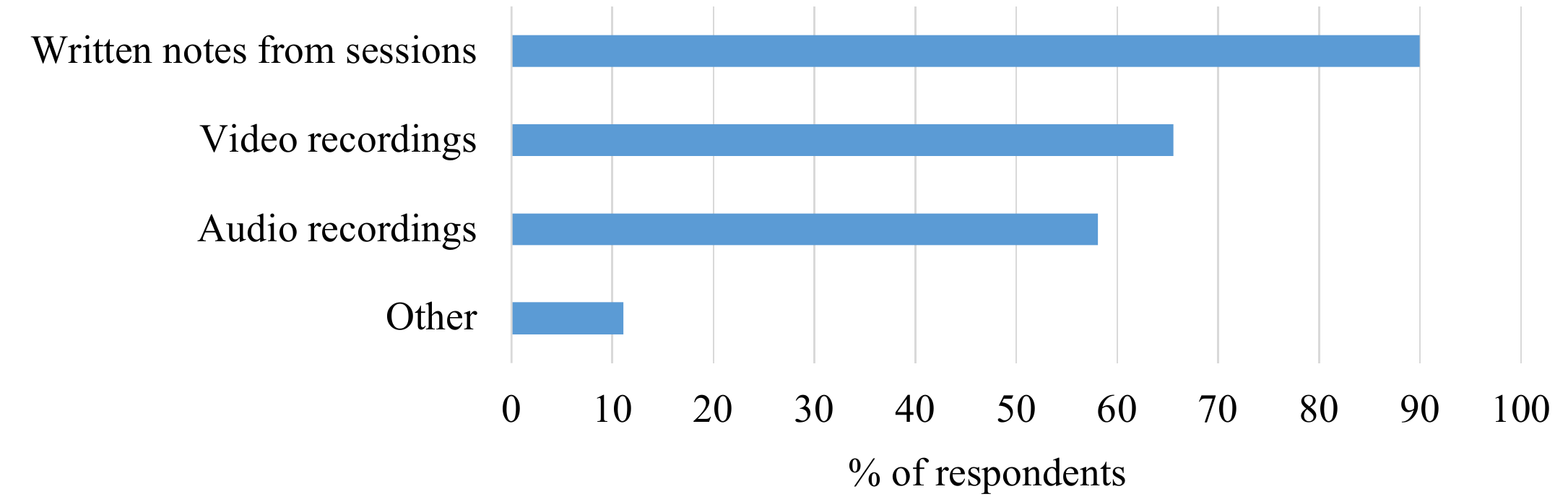}
  \caption{Resources reviewed when conducting data analysis $(N=279, multi-select)$}
  \Description{Bar graph of resources reviewed when conducting analysis where 90\% of respondents used written notes}
  \label{fig:Resources}
\end{figure}

Figure \ref{fig:Resources} shows that the majority (90\%) of the respondents reviewed written notes from sessions (either their own or their team members' observational notes), 66\% of them reviewed video recordings, and 58\% of them reviewed audio recordings.  
11\% of the respondents reported other resources including automated transcripts, qualitative data from interviews, testing software (e.g., UserTesting \cite{usertesting_usertesting_2020}, UserZoom \cite{userzoom_userzoom_2021}), heatmaps and polls, and periodic industry reports.

This is in line with the findings of a 2020 international survey showing that 89\% of respondents reviewed observation notes and 77\% reviewed session recordings (both audio and video) \cite{fan_practices_2020}. 
However, an earlier interview study from 2010 found that the majority of respondents relied mainly on their notes and memory for analysis instead of doing a full analysis of recordings \cite{folstad_analysis_2010}.                                          
This suggests that more UX practitioners use recordings more often nowadays than before, which could be due to recent advancements in technology and remote usability testing. 
However, written notes remain the most used method, highlighting the importance in the quality of the original notes from sessions. 
 
\begin{figure}[h]
  \centering
  \includegraphics[width=\linewidth]{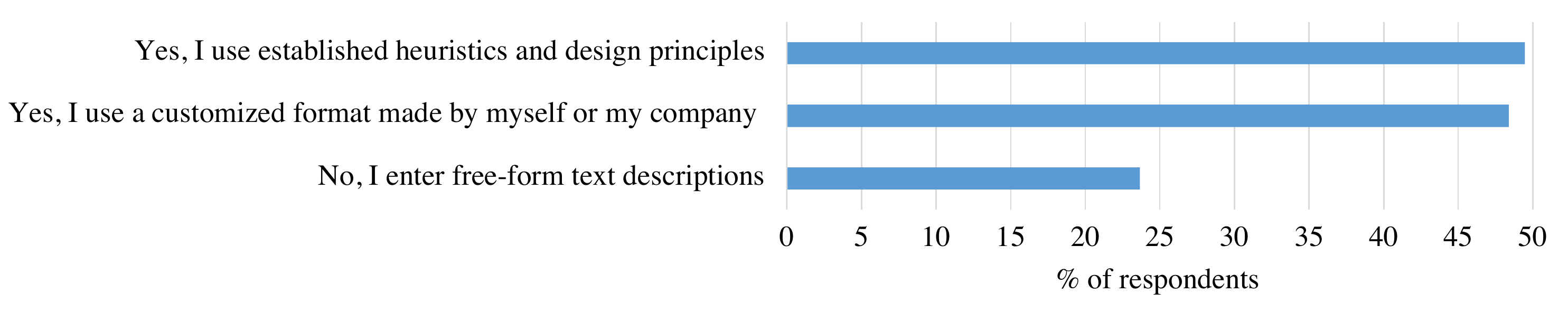}
  \caption{Structured formats used when describing identified problems $(N=279, multi-select)$}
  \Description{Bar graph of structures used when describing identified problems where around half of the respondents used established heuristics and half used customized heuristics}
  \label{fig:IfStructure}
\end{figure}

Figure \ref{fig:IfStructure} shows that almost half (49\%) of the respondents used established heuristics and design principles. Similarly, almost half (48\%) followed a customized format. Since respondents were allowed to select multiple options, they could have used a combination of established and customized formats as well. 

\begin{figure}[h]
  \centering
  \includegraphics[width=\w\linewidth]{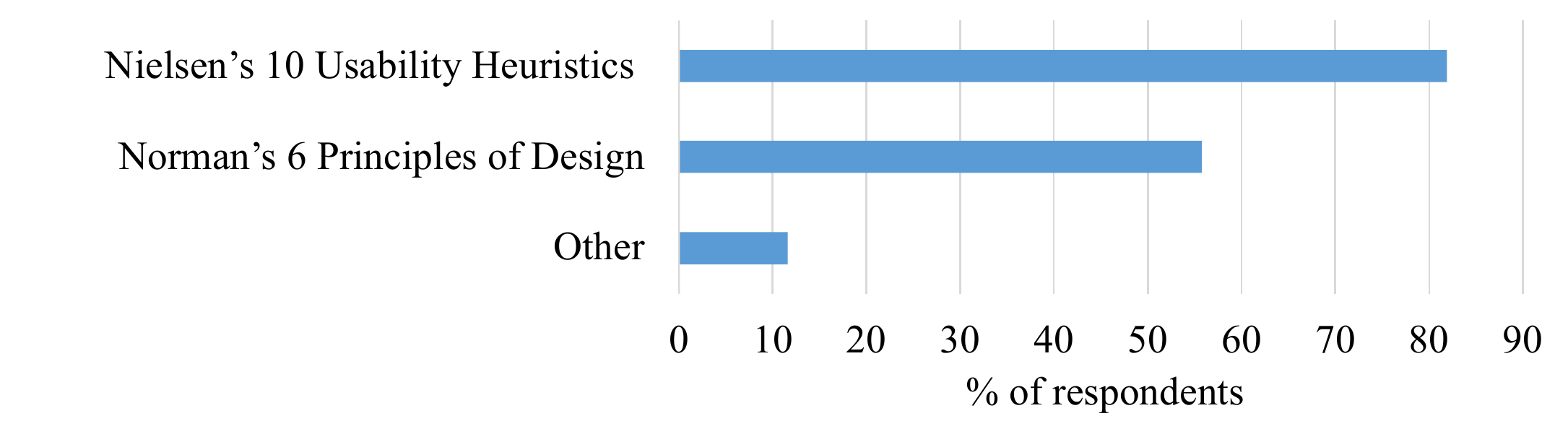}
  \caption{Usage of established heuristics or design principles $(N=138, multi-select)$}
  \Description{Bar graph of usage of established heuristics or design principles with 82\% of respondents using Nielsen's 10 Usability Heuristics and 56\% using Norman's 6 Design Principles}
  \label{fig:Heuristics}
\end{figure}

Specifically, as shown in Figure \ref{fig:Heuristics}, Nielsen's 10 usability heuristics \cite{nielsen_10_1994} were the most popular among the respondents with the majority (82\%) of them using it when detecting and describing problems. 
Over half (56\%) of them used Norman’s 6 design principles \cite{norman_design_2002}.  
Furthermore, respondents also reported using other UX evaluation guidelines and standards, including the International Usability and UX Qualification Board \cite{uxqb_international_2021}, the MeCUE questionnaire for UX assessment \cite{minge_mecue_2018}, the Gestalt Principles of Design \cite{koffka_principles_2013}, heuristics for mobile computing \cite{bertini_appropriating_2006-1}, and the ISO 9241-110 Standard for Ergonomics of Human-System Interaction \cite{iso_iso_2020}. 

When using these standard heuristics, principles, and guidelines, respondents also modified them to cater to their companies' needs. The free responses revealed their reasons, which included adding more context to the heuristics to make it clearer to non-experts. Many of them did not change the core meaning of the heuristics but rather the language, which allowed them to better communicate the findings to stakeholders and other audiences.

\begin{figure}[h]
  \centering
  \includegraphics[width=\w\linewidth]{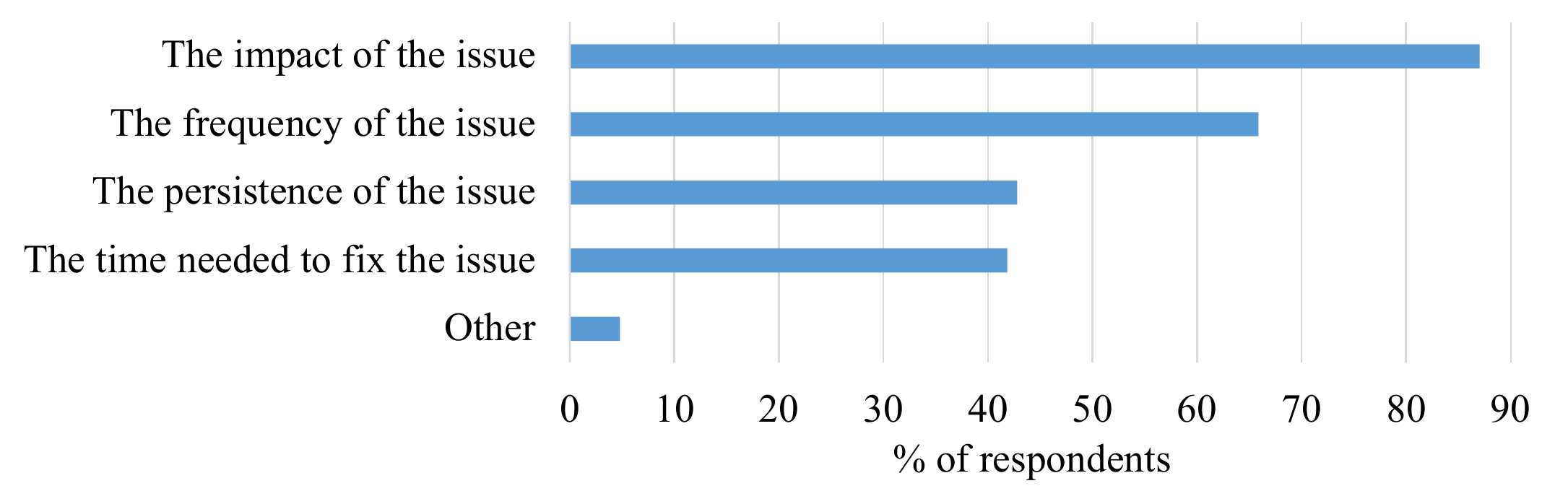}
  \caption{Factors considered when assigning severity levels $(N=208, multi-select)$}
  \Description{Bar graph of factors considered when assigning severity levels where 87\% of respondents selected the impact of the issue}
  \label{fig:FactorsofSeverity}
\end{figure}

In addition to identifying UX problems, most (75\%) of the respondents rated the severity of the problems.
Since assigning severity classifications was suggested by prior work \cite{folstad_analysis_2010}, our findings showed the different factors that respondents considered while doing so (see Figure \ref{fig:FactorsofSeverity}). 
Specifically, the majority (87\%) of them considered \textit{``the impact of the issue''}, followed by \textit{``the frequency of the issue''} (66\%), \textit{``the persistence of the issue''} (43\%), and \textit{``the time needed to fix the issue''} (42\%). Moreover, a small percentage of respondents (5\%) reported also considering the potential cost to fix the issue, the scope of the issue, and whether it had any relation to previously uncovered problems. 

\subsection{RQ2: Collaborative Data Analysis Practices, Challenges, and Desired Improvements}

\subsubsection{Collaboration Purposes and Circumstances} 
\begin{figure}[h]
  \centering
  \includegraphics[width=\w\linewidth]{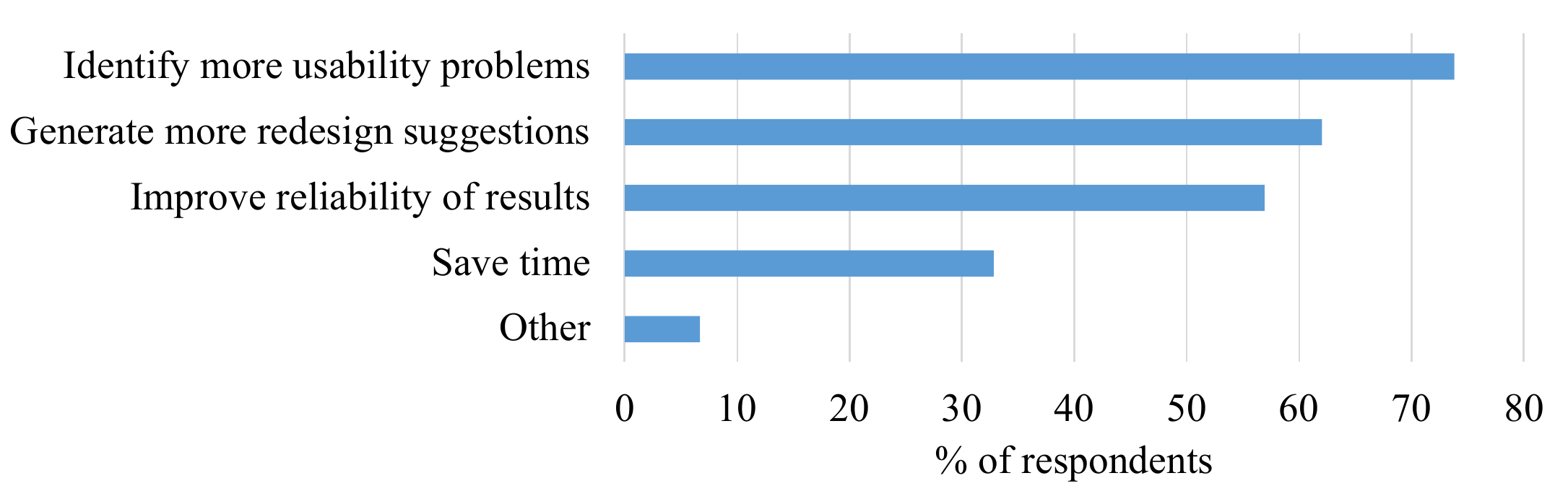}
  \caption{Purposes of collaboration between UX practitioners $(N=195, multi-select)$}
  \Description{Bar graph of the purposes of collaboration between UX practitioners where 74\% selected to identify more usability problems}
  \label{fig:Purpose}
\end{figure}

The top three purposes for collaboration, as shown in Figure \ref{fig:Purpose}, were to \textit{identify more usability problems (74\%)}, \textit{generate more redesign suggestions (62\%)}, and \textit{improve reliability of results (47\%)}. This finding suggests that respondents were aware that collaboration could increase the completeness and reliability of their analysis, which would be beneficial to address the \textit{``evaluator effect''} \cite{hertzum_evaluator_2001}. 

In this survey, \textit{improving reliability of results} was only the third-frequently chosen option which is in contrast to prior findings of this being the top driver for collaboration \cite{folstad_analysis_2012}. 
One explanation for this could stem from the format of the question. In the prior survey, it was a single-choice question where 47\% of respondents selected \textit{improving reliability of results}. 
In this survey, it was selected by over half (57\%) of the respondents. This result suggests that most of respondents did pay attention to the reliability of their analysis. However, it was less frequently considered as compared to more result-oriented purposes such as increasing the quantity of usability problems and redesign suggestions. 
Additionally, respondents provided free-form responses in terms of their purposes for collaboration, and the top responses were to ``save money'', ``consolidate paper work'', and ``ensure everyone's voices are heard.''
The awareness of the importance of collaboration was demonstrated by respondents who mentioned that \textit{different perspectives are critical to identifying as many issues and potential solutions as possible}. 

\begin{figure}[h]
  \centering
  \includegraphics[width=\linewidth]{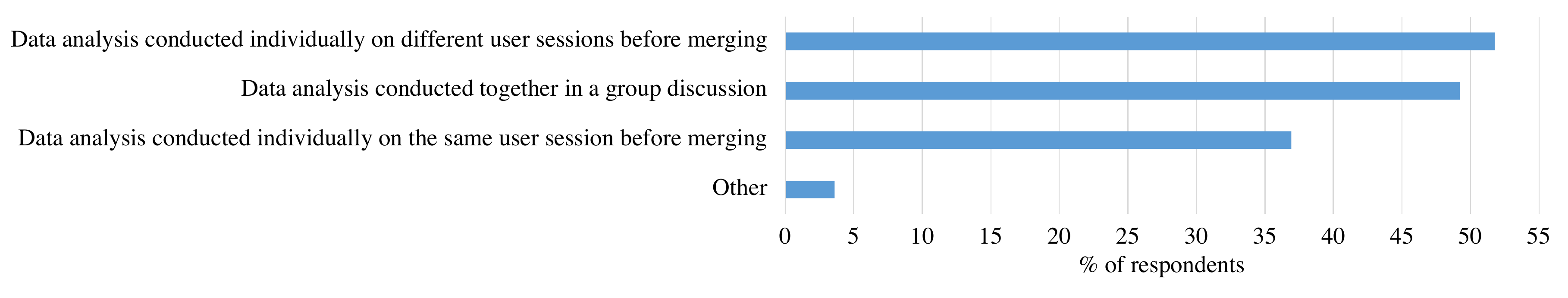}
  \caption{Circumstances of collaboration between UX practitioners $(N=195, multi-select)$}
  \Description{Bar graph of the circumstances of collaboration between UX practitioners where 52\% of respondents conducted data analysis individually on different user sessions before merging}
  \label{fig:Circumstance}
\end{figure}

As shown in Figure \ref{fig:Circumstance}, most of the collaboration between colleagues happened when they conducted data analysis \textit{individually on different user sessions before merging} (52\%). 
This resembles a \textit{divide and conquer} approach and allows the UX practitioners to save time as it reduces the workload for a single practitioner. The second most common approach was to conduct analysis \textit{together in a group discussion} (49\%).
Furthermore, over half (52\%) of these discussions took place \textit{directly after} the usability testing session, 46\% took place \textit{within a week}, and only 2\% took place \textit{more than a week after}. 
This is in line with our earlier finding that the majority (70\%) of respondents only have less than a week to analyze data from usability testing (see Figure~\ref{fig:TimeCost}).

In contrast, only 37\% of the respondents conducted data analysis \textit{individually on the same user session before merging}. 
This phenomenon was echoed in the free-form responses as many respondents mentioned that even though they worked on a team with other UX practitioners, only one person was responsible for analyzing all usability testing sessions for a single product in the company.

Another vital factor to successful collaboration is having effective communication between colleagues \cite{stern_sharing_2008-1,jones_determinants_2019}. 
Thus, we asked respondents what type of communication they use most often to discuss work-related content with their colleagues. 
Almost three-quarters (73\%) of the respondents chose synchronous methods such as meetings or phone calls, whereas 27\% use asynchronous methods such as emails. 
Indeed, their choice of communication method is in line with their preference as the majority (68\%) selected ``somewhat agree'' or ``strongly agree'' to the prompt \textit{``I prefer to talk to my colleague directly rather than type through chat''} (Md=4, IQR=2). 
In their explanations, respondents mentioned that \textit{``there is always a need for synchronous communication in between asynchronous discussions''} and that \textit{``it's hard to put together a presentation or align on findings without real-time discussions.''}

\subsubsection{Platforms that Support Collaboration} 

\begin{figure}[h]
  \centering
  \includegraphics[width=\w\linewidth]{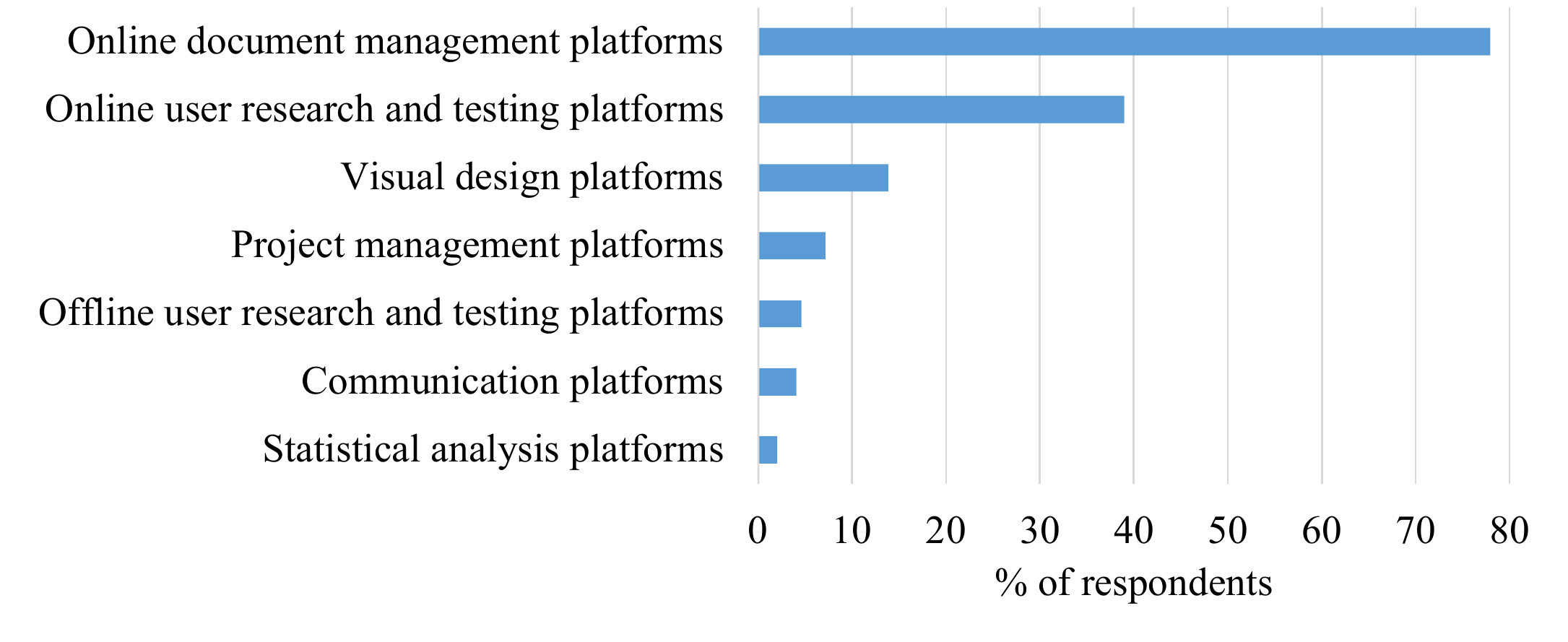}
  \caption{Different categories of platforms used for collaboration while conducting data analysis $(N=195, multi-select)$}
  \Description{Bar graph of different platforms used for collaboration while conducting data analysis where the majority (78\%) of respondents used online document management platforms}
  \label{fig:Tools}
\end{figure}

The respondents used a total of 31 different platforms for collaboration while conducting data analysis of user sessions, which were grouped into various categories (see Figure \ref{fig:Tools}).

The majority (78\%) of respondents used \textit{online document management platforms} which included Google Workspace (e.g., Docs, Sheets, Slides, and Forms) \cite{google_google_2021-1} and Microsoft SharePoint \cite{microsoft_microsoft_2021-1}. 
The next popular category was \textit{online user research and testing tools}, which was used by 39\% of the respondents. Examples include UserTesting \cite{usertesting_usertesting_2021}, UserZoom \cite{userzoom_userzoom_2021}, Useberry \cite{useberry_useberry_2021}. 
These platforms provide functionalities like note-taking and marking events on the video progress bar on top of basic usability test support such as screen recording, survey administration, and results exporting.
Respondents often used combinations of these top two categories. For example, one respondent described that they used Google Slides and Sheets to collaborate during data analysis since it was fast and real-time, but still found it easier to share a link to the UserTesting study when reviewing videos and participant data.

Although \textit{offline user research and testing tools} offer similar functionalities, they do not support flexible collaboration. For example, it is difficult for multiple UX practitioners to gain access to the same session recordings when stored locally. In fact, only 5\% of the respondents currently used offline platforms, such as Morae \cite{techsmith_morae_2020}, which will likely continue to decrease as online remote user testing platforms gain more popularity~\cite{moran_remote_2020}. 
Other categories include \textit{visual design platforms} such as Miro \cite{miro_miro_2021}, Mural \cite{tactivos_inc_mural_2020}, and Figma \cite{figma_figma_2021} (14\%), \textit{project management platforms} such as Jira \cite{atlassian_jira_2021}, Confluence \cite{atlassian_confluence_2021}, and Trello \cite{atlassian_trello_2021} (7\%), and \textit{communication platforms} such as Microsoft Teams \cite{microsoft_microsoft_2021} and Zoom \cite{zoom_video_2021} (4\%). 

Respondents had mixed feelings about the use of many tools to support their collaborative data analysis. On the one hand, respondents mentioned that they like having the option of many tools as it gives them the choice to use something that they are comfortable with. On the other hand, respondents felt that \textit{``if there are too many tools, it can get confusing especially if there isn't a main one the team has agreed to rely on most.''}

\subsubsection{Challenges of Collaboration}

\begin{figure}[h]
  \centering
  \includegraphics[width=\linewidth]{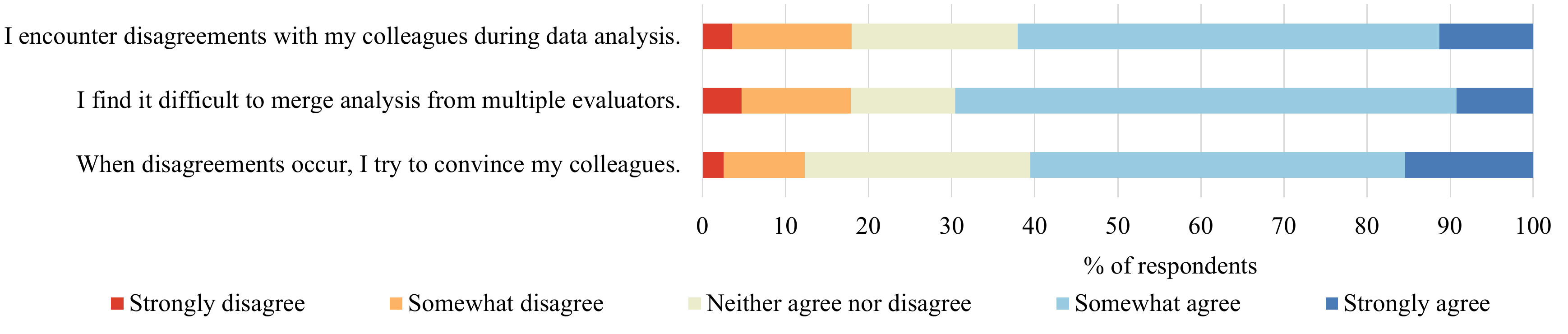}
  \caption{Likert scale rating of challenges encountered when conducting analysis $(N=195)$}
  \Description{Distribution of likert scale ratings for challenges during analysis}
  \label{fig:Challenges}
\end{figure}

As shown in Figure~\ref{fig:Challenges}\footnote{This divergent color scheme was based on an established color-blind friendly palette \cite{toh_colour_2021}.}, respondents encountered various challenges during collaboration. 
Almost two-thirds (62\%) of respondents encountered situations where they disagreed with their colleagues during data analysis (Md=4, IQR=1). 
Moreover, finding it difficult to merge analysis from multiple evaluators was also considered as a challenge by over two-thirds (70\%) of respondents (Md=4, IQR=1). 
When disagreements occurred, 61\% of them would make a case for their decisions and try to convince their colleagues (Md=4, IQR=1). 

Regarding their approaches to managing disagreements, most of the respondents (82\%) would \textit{have a discussion in-person or through video/voice call} and 33\% would \textit{have a discussion through chat}. 
Other strategies include \textit{involving another person as the mediator} (24\%) and \textit{asking another person to be the tie-breaker} (23\%). 
Some respondents provided additional strategies in the free responses, which includes \textit{approaching management or a supervisor with both sets of findings and asking them to make the decision}.
One respondent stated that they rarely disagreed on a usability problem, but rather the priority of a problem. When this was the case, they \textit{``refer[red] to the main goal of the study and then reflect[ed] on what the client would find most helpful and actionable.''} 
Another response was creating an anonymous board to list down the issues and then voting on it at the next group meeting. 
In the case where a designer was part of the discussion, one respondent mentioned they would let the designer of the test product make the final call. 
Interestingly, one respondent wrote \textit{``have manager make decision to go with my advice over the others''}, which suggests that they would try to convince their manager to listen to their advice first. 
If they are under time pressure and still can't reach an agreement on a certain issue, respondents mentioned two strategies in particular: (1) asking the leader to make the final decision, and (2) presenting their results with an explanation that covers both interpretations since \textit{``most likely one person isn't wrong, it is simply that the interpretations are based on different viewpoints.''}

\begin{table*}[htb!]
  \caption{Analysis and collaboration practices that have significant differences for varying \textit{years of UX experience}}
  \label{tab:experienceCorrelations}
  \begin{tabular}{lr}
    \hline
    \rowcolor[gray]{0.9}
    Collaboration practices & \\
    \hline
    Found it time consuming to conduct data analysis & $\chi^2(16)=32.1, N=279, p<.01, V=.17**$ \\
    Used a customized problem description format & $\chi^2(4)=15.2, N=279, p<.01, V=.23**$ \\
    Collaboration purpose is to improve reliability of results & $\chi^2(4)=11.6, N=279, p<.05, V=.20*$ \\
    \hline
\end{tabular}
\end{table*}
The responses for the greatest challenges that UX practitioners have encountered when collaborating on data analysis were analyzed and grouped into the following themes: 

\textbf{Data-related challenges:} 
\begin{enumerate}[nosep]
    \item \textit{Data storage and access issues}: Respondents found it difficult to store all the data and analysis associated with usability testing in \textit{``a visually organized, quick and easily accessible way for follow-up discussions''} with colleagues.
    \item \textit{Data analysis challenges}: Respondents mentioned some aspects of data analysis were especially tedious or difficult, such as \textit{``clean[ing] up software produced transcripts''}, \textit{``conduct[ing] qualitative analysis of text''}, \textit{``setting the frame for insight extraction''}, \textit{``synthesizing findings into actionable recommendations''}, and \textit{``remember[ing] a specific point of time after the session has ended''}. 
\end{enumerate}


\textbf{Coworking and communication challenges:}
\begin{enumerate}[nosep]
    \item \textit{Inconsistent coding strategies}: Respondents had difficulty reaching agreement on what codes or labels the transcript should be used to analyze texts. Respondents felt that \textit{``it's not always clear how to analyze and it's hard to change in the middle.''} \mc{It was also} difficult to \textit{``merge data analysis from many researchers and notify others about the updates''}.
    \item \textit{Contrasting priorities}: Respondents struggle to convince colleagues to have a \textit{``problem-first'' mentality} instead of rushing to a solution without careful research and analysis. Another respondent even had difficulty \textit{``getting everyone in the room to listen to the analysis due to different priorities and values between colleagues and clients''}.
    \item \textit{Mismatch of experience}: Respondents were not confident in their colleagues' capabilities to analyze data. The team consisted of varying levels of \textit{``maturity when it comes to UX understanding''}, leading to inefficiencies in collaboration such as \textit{extra time [needed] to set the frame for analysis}. 
    \item \textit{Biased feedback}: Respondents found that getting honest opinions from their team was sometimes a challenge as their colleagues were \textit{``biased to give positive feedback''}.
\end{enumerate}


\textbf{Environmental challenges:} 
\begin{enumerate}[nosep]
    \item \textit{Lack of resources}: Many respondents did not have enough time or budget for multiple people analyze each recording. One of them even needed to \textit{``convince stakeholders that it requires more time than they think''} and another mentioned that there was \textit{``no budget for real analysis''}. We also found that 30\% of the respondents were the only UX practitioner on their team and one mentioned \textit{``lack of a second opinion''} as their biggest challenge. This illustrates the prevalence of resource constraints, where they have to work alone without others to double check their analysis results.
    \item \textit{Remote work issues}: Respondents mentioned that \textit{``communicating with the team is challenging when we can't speak''} due to working from home. The process of getting quick feedback on a usability problem from the team requires waiting for a response or setting up a meeting, which causes delays and inefficiencies. \rv{The COVID-19 pandemic---the time at which this survey was conducted---might affect this finding as many workplaces shifted from in-office to working from home \cite{brynjolfsson_covid-19_2020}. We will discuss it in Section~\ref{sec:limitations}.} 
    \item \textit{Limitations of software tools}: Some respondents were frustrated with managing \textit{``so many columns in the Excel sheet and it's too tiring to go back and forth [between columns]''}. Others were used to paper analysis where they can \textit{``put multiple different color post-its or highlight the sheets with notes''}. However, these features are missing in some analysis software and they feel the need for a \textit{``quick way to relate different types of data''}.
\end{enumerate}


\subsubsection{Desired Features in New Collaboration Tools}

\begin{figure}[htb!]
  \centering
  \includegraphics[width=\linewidth]{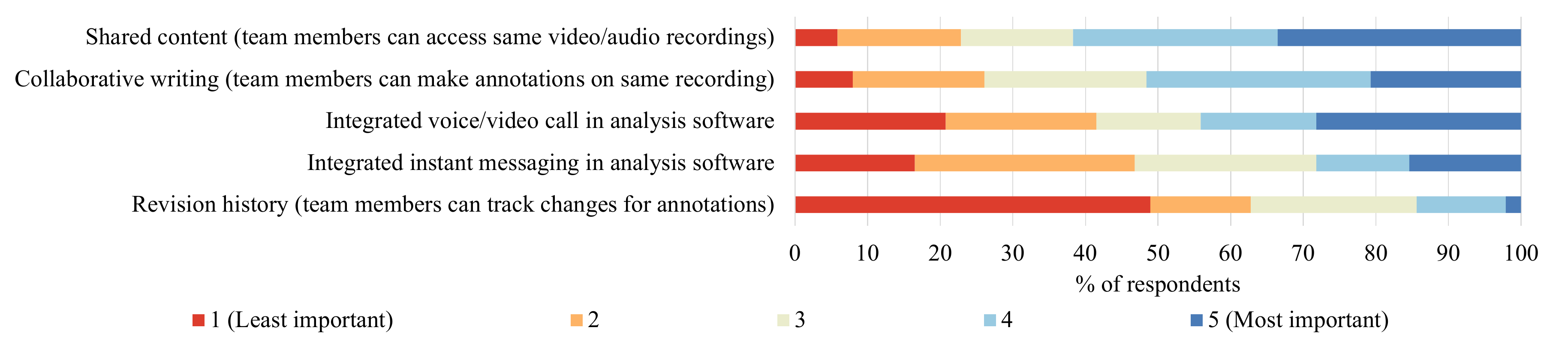}
  \caption{Rating of desired collaboration features in the data analysis tool $(N=195)$}
  \Description{Distribution of ratings for of desired collaboration features in new collaborative analysis tool}
  \label{fig:RankFeatures}
\end{figure}

We also wanted to understand what features respondents would prefer in an all-inclusive user testing analysis tool that supports collaboration between UX practitioners. 
They assigned each of the five collaboration features a value from 1 (least important) to 5 (most important), shown in Figure \ref{fig:RankFeatures}. 

From the most important to the least, the features were 1) \textit{shared content (team members can access same video/audio recordings)} (Md=4, IQR=2), 
2) \textit{collaborative writing (team members can make annotations on the same recording)} (Md=4, IQR=2), 
3) \textit{integrated voice/video call in analysis software} (Md=3, IQR=3), 
4) \textit{Integrated instant messaging in analysis software} (Md=3, IQR=2), 
and 5) \textit{Revision history (team members can track changes for annotations)} (Md=2, IQR=2). 

Overall, more than half (58\%) of the respondents ``somewhat agreed'' or ``strongly agreed'' that they needed new tools to support their data analysis and collaboration process (Md=4, IQR=1). 
The main suggestions for improvements gathered from the free-form responses include: 
\begin{itemize}[nosep]
    \item Being able to schedule meetings and attach \mc{materials} in the invitation to team members from within the data analysis tool \mc{so that the materials are} readily available and organized.
    \item Having an easier way to export findings and notes from the data collection tool to a usable format like Word.    
    \item Having one spot to consolidate the findings with coworkers in addition to the notes and session recordings.
\end{itemize}

\subsection{RQ3: Correlations} 
\label{sec:correlations}




\mc{We} conducted Pearson's chi-squared test for sets of categorical data to evaluate how likely it is that any observed differences arose by chance \cite{cohen_statistical_1988}. We also utilized Cramer's V as a measure of the strength of the association between variables (effect size) \cite{cohen_statistical_1988}. 

\begin{table*}[htb!]
  \caption{Analysis and collaboration practices that have significant differences between people on varying \textit{sizes of UX teams}}
  \label{tab:teamCorrelations}
  \begin{tabular}{lr}
    \hline
    \rowcolor[gray]{0.9}
    Collaboration practices & \\
    \hline
    Used a customized problem description format & $\chi^2(2)=8.3, N=195, p<.05, V=.15*$ \\
    Included severity ratings of usability problems & $\chi^2(2)=7.6, N=195, p<.05, V=.14*$ \\
    Managed disagreements by having a mediator & $\chi^2(2)=7.3, N=195, p<.05, V=.14*$ \\
    Tried to convince colleagues when in disagreement & $\chi^2(8)=22.2, N=195, p<.01, V=.12**$\\
    \hline
\end{tabular}
\end{table*}

\subsubsection{Years of UX Experience}

\begin{figure}[h]
  \centering
  \includegraphics[width=\w\linewidth]{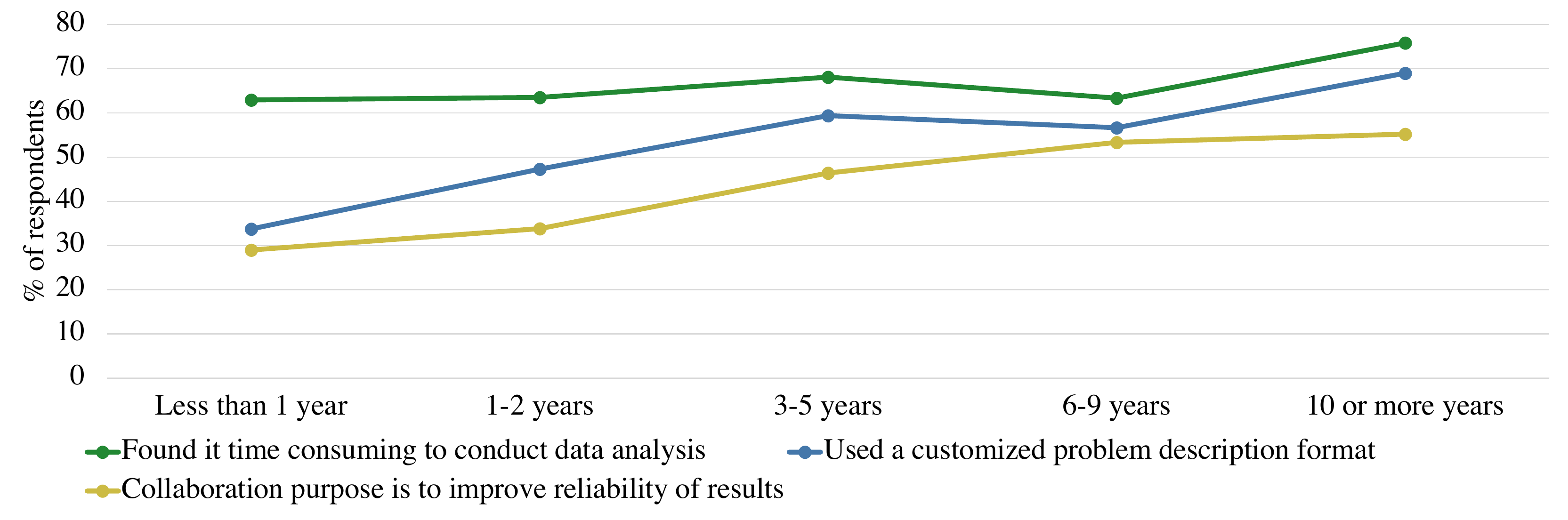}
  \caption{Line chart showing three significant correlations according to years of UX experience.}
  \Description{Line chart showing three significant correlations according to team size.}
  \label{fig:YearsExperienceCorrelations}
\end{figure}

Table ~\ref{tab:experienceCorrelations} shows the collaboration practices that have significant differences between respondents with different years of UX experience\footnote{significant levels: *: p<.05, **: p<.01, ***: p<.001}. 
Fig \ref{fig:YearsExperienceCorrelations} provides a visualization of these observed trends. 

First, as years of UX experience grew, respondents were more likely to agree and strongly agree that \textit{it is time-consuming to conduct data analysis}. Although overall 66\% of respondents felt it was time-consuming, the highest proportion (76\%) occurred in respondents with over 10 years. One reason may be that experienced respondents have a better understanding of the workload and difficulty in conducting a thorough data analysis based on their years of experience; it could also be that they tend to have other responsibilities that take up their time such as managing projects and meeting with stakeholders.

Second, the more years of UX experience that respondents had, the more likely they might \textit{use a customized format} made by themselves or their companies when composing analysis reports. Almost 70\% of respondents with 10 or more years of UX experience used a custom format, compared to only 34\% with less than a year. Practitioners with more experience tended to modify heuristics to meet specific needs while novice practitioners followed pre-established formats and guidelines such as Nielsen's Heuristics \cite{nielsen_10_1994}.


Lastly, respondents with greater years of UX experience were more likely to treat \textit{``improve reliability of results'' as a purpose for collaboration} since over half (55\%) of respondents with 10 or more years of experience and 6-9 years (53\%) selected it compared to just 29\% of respondents with less than 1 year and 34\% of 1-2 years. This may be because more experienced practitioners were more aware of the importance of reliability in the analysis results.

\subsubsection{Size of UX Team}

\begin{figure}[h]
  \centering
  \includegraphics[width=\w\linewidth]{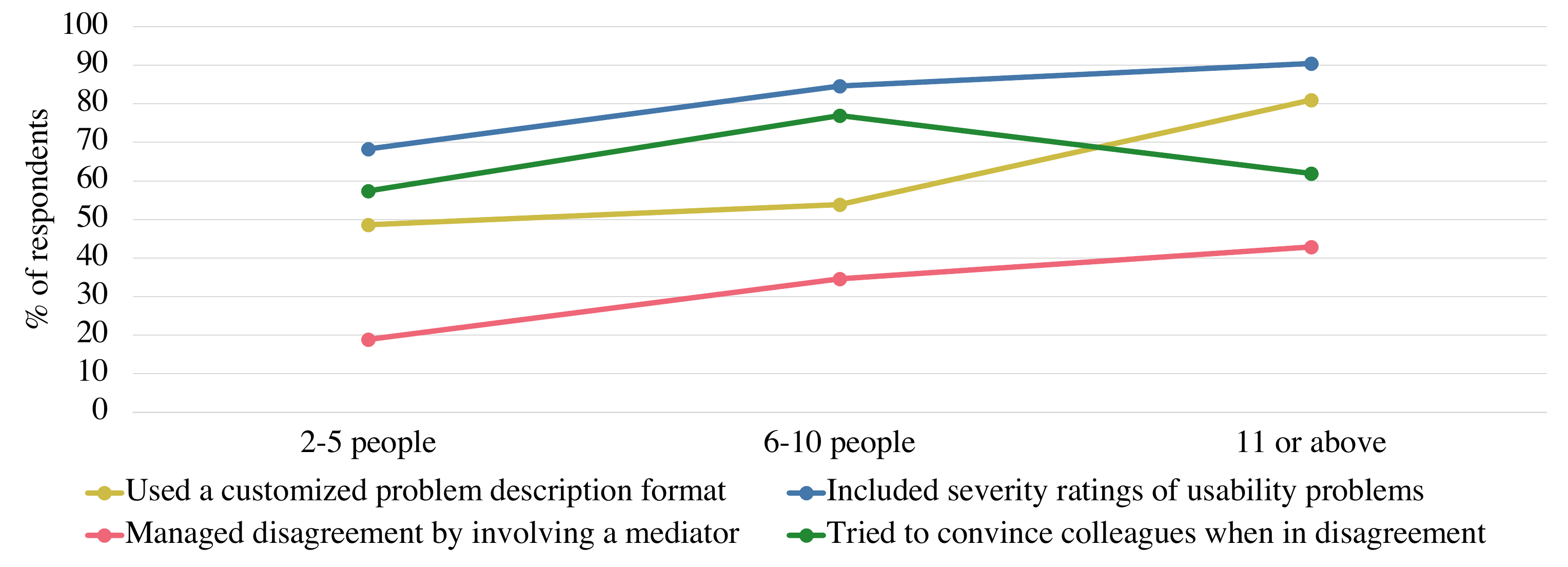}
  \caption{Line chart showing four significant correlations according to team size.}
  \Description{Line chart showing four significant correlations according to team size.}
  \label{fig:TeamSizeCorrelations}
\end{figure}

Table ~\ref{tab:teamCorrelations} shows the collaboration practices that have significant differences between respondents on different size UX teams, which is visualized in Fig \ref{fig:TeamSizeCorrelations}.

First, respondents on a larger team were more likely to \textit{use a customized format made by themselves or their companies}. The majority of respondents on a team of 11 and above used a self-defined customized format (81\%), about half of respondents on a team of 6-10 used it (54\%), and less than half of respondents on a team of 2-5 used it (49\%). 

Similarly, respondents on a larger team were more likely to \textit{include ratings of severity for the problems identified}. 91\% of respondents on a team of 11 and above included it, whereas only 68\% of them on a team of 2-5 people did. 
These two trends may be because larger teams tend to have more resources to create custom heuristics or modify existing ones to fit their needs and prioritize the severity of the identified problems.

Another trend seen in Fig \ref{fig:TeamSizeCorrelations} is that respondents on a larger team were significantly more likely to \textit{involve another person as the mediator when managing disagreements between colleagues}. 43\% of respondents on a team of 11 or more people would do so, compared to just 35\% of respondents on a team of 6-10 people and 19\% of respondents on a team of 2-5 people. This may be because larger teams tend to have more people available to play the role of a mediator. For teams with only two people, involving another person takes more effort since they might need to find someone from another team and explain the context on the disagreement.

Lastly, respondents on a team of 6-10 people (77\%) were more likely to \textit{make a case for their decisions and try to convince their colleague when disagreements occur} compared to 57\% of them on a team of 2-5 people and 62\% of them with 11 and above. 

\section{Discussion}



In this section, we discuss the key findings and present related design considerations for creating faster-paced and more reliable analysis and collaboration among UX practitioners. 

\subsection{Modes of Collaboration}

For the respondents whose companies have the resources for multiple UX practitioners, collaboration generally falls under three modes: the most \mc{frequent mode used} by over half (52\%) of the respondents is to \textit{independently analyze different portions of the data first with little or no overlap before collaboration (i.e., divide and conquer)}. 
This mode highlights how UX practitioners cope with the tension between reliability and efficiency of data analysis \cite{fan_practices_2020}. 
However, while evaluators may collectively find more problems, each usability test session is essentially only reviewed by one evaluator, who might miss problems due to the ``evaluator effect'' \cite{hertzum_evaluator_2001}. 

The second-most frequently used (49\%) mode is to \textit{collaboratively analyze the sessions with little or no independent analysis (i.e., group analysis)}. 
The prevalence of this type of group analysis echoes the findings of F{\o}lstad et al \cite{folstad_analysis_2012}.
However, its advantages and disadvantages are not well-understood as there is a gap in the literature on conducting analysis in groups \cite{folstad_analysis_2012}. 
Law and Hvannberg studied the process of consolidating usability problems individually vs. collaboratively in a within-subjects design and revealed a tendency towards collaborative settings \cite{law_consolidating_2008}.
However, a follow-up between-subjects study found no significant differences between practitioners in the two settings, suggesting that the consolidation process does not benefit from positive group decision effects \cite{hoffmann_consolidation_2019}. 
It is worth noting that participants from both studies were novice practitioners (i.e. students with limited or no background knowledge \mc{in} UX), so their results might not be generalizable to practitioners with more experience. Indeed, our results (Table~\ref{tab:experienceCorrelations}) also suggest that \mc{years of UX experience} may affect their collaboration practices.
As a result, the benefits of group analysis remain unclear for UX practitioners who have a variety of experience. 
In order to preserve the reliability of the results, it is critical to investigate best-practice guidelines and procedures for group analysis. 

The third mode is to \textit{independently analyze the same set of data and then collaborate}. 
Although prior research suggested that this type of collaboration could improve reliability \cite{hertzum_evaluator_2001} and is recommended \cite{folstad_analysis_2012}, our study found that only 37\% of respondents practiced this collaboration mode. One potential reason might be that it requires more human effort for the same amount of data as it requires more than one practitioner to look at the same session. Furthermore, these practitioners would have to spend additional time consolidating their analyses.  
Indeed, 70\% of our respondents felt it was difficult to merge analysis from individual practitioners, which may lead to more time costs. 

\subsubsection{Design Consideration Based on Collaboration Modes}

\textbf{Establish processes for analysis in groups and balance the trade-offs between efficiency and validity.} We uncovered the three modes of collaboration among UX practitioners. However, it remains an open question of what are the advantages and disadvantages of the three collaboration modes in terms of improving completeness and reliability of usability testing results.
Although the third mode was recommended by prior work \cite{folstad_analysis_2012}, such mode has higher time costs compared to the other two modes. 
The other modes are at odds with recommended practices where evaluation performance should be based on thoroughness, validity, and reliability \cite{folstad_analysis_2010, hartson_criteria_2003} with the aim of minimizing the \textit{``evaluator effect''} \cite{hertzum_evaluator_2001}.
However, in reality, the time and resources of a particular company should also be taken into account. 
As previously mentioned, the lack of time and resources is a key challenge that respondents faced, which was echoed by previous surveys of UX practitioners in industry \cite{norgaard_what_2006, chilana_understanding_2010, folstad_analysis_2012, fan_practices_2020}. 
Thus, one possible advantage for group analysis could be to save time as more people may lead to faster identification of usability problems and having in-situ discussions could avoid the need of a second meeting to merge the results from different practitioners.
Due to the trade-offs between time and reliability, one important question is to help UX practitioners determine when they have conducted sufficient analysis (e.g. at what point can they stop). 
For example, Nielsen and Landauer presented a model showing the proportion of usability problems in an interface found using various numbers of practitioners \cite{nielsen_mathematical_1993}. 
The proportion increased as the number of practitioners increased, thus the use of at least three practitioners was recommended \cite{nielsen_mathematical_1993}. 
However, this model was based on independent analysis of the same data from multiple practitioners. Thus, it would be interesting to investigate methods to extend this model for analysis conducted in groups.
Future research should explore ways of collaboration that achieve a balance between validity and efficiency in UX data analysis, which is a key challenge that UX practitioners face \cite{fan_practices_2020}.

\subsection{Tools Used for Analysis}

Our findings suggest that there is a lack of resources in the types of tools available to support both data analysis and collaboration at the same time.
The 31 different tools used by the respondents fell into seven categories (e.g. \textit{online document management platforms}, \textit{online user research and testing tools}, and \textit{visual design platforms}). 
The variety in the types of tools provides evidence that respondents have to switch between different platforms. 
This may cause overhead in trying to review data, conduct analysis, and then communicate and merge these findings with colleagues. 
During this process, there could be many difficulties such as storing and keeping track of separate recordings and analysis documents. 
These \textit{data storage and access issues} were mentioned as a key challenge by respondents.
Respondents also had mixed feelings about the use of many tools. Despite being flexible, having multiple tools was also confusing. 

\subsubsection{Design Consideration Based on Analysis Tools}

\textbf{Develop \textit{one integrated environment} to support UX practitioners in reviewing usability test sessions, performing data analysis, and collaborating.}
One of the main improvements that respondents suggested was \textit{``having one spot to consolidate the findings with coworkers.''}
Building a platform that allows UX practitioners to review usability test sessions, conduct data analysis, and communicate with their team would also mitigate the overhead of needing to use multiple tools. 
Numerous commercial tools have been developed to support UX practitioners with conducting usability test and reviewing test session data, including UserZoom~\cite{userzoom_userzoom_2021} and FullStory~\cite{fullstory_fullstory_2021}. 
While these tools allow multiple team members to access the same recordings, their data analysis capabilities are mostly limited to session playback, note-taking, and mouse point clouds. 
Thus, it is important to design tools to meet the unique analysis needs that generic tools (e.g., Google Sheets \cite{google_google_2021-1}) do not support well, such as adding heuristics, which was a common practice among 76\% of the respondents and have been shown to be effective in enabling practitioners to uncover more usability problems \cite{law_analysis_2004}. In addition, such tools should consider to visualize subtle behavioral patterns that might indicate UX problems~\cite{fan_concurrent_2019,Fan2021OlderAdults}, such as the sentiment of users' words and their abnormal tones. What's more, such tools should also integrate collaboration features (e.g. dimension coverage of data \cite{sarvghad_exploiting_2015, zhao_supporting_2018}, radar views \cite{gutwin_descriptive_2002}, synchronized annotation timeline \cite{satybaldiev_coat_2019}) to support UX practitioners to discuss and consolidate their individual analyses while still having access to both the raw session recordings and their analysis notes in one integrated environment. 
For example, collaborative data analysis can benefit from \textit{displaying data dimension coverage of history}~\cite{sarvghad_exploiting_2015,sarvghad_visualizing_2017} or highlights of previously investigated data in a graph visualization~\cite{zhao_supporting_2018}.
In addition, radar views provide awareness of where
one is working relative to their remote collaborators in a
virtually shared workspace \cite{gutwin_descriptive_2002}.
A synchronized annotation timeline supports collaborative video annotations by distributing work and sharing results simultaneously \cite{satybaldiev_coat_2019}. 
Collaborative writing tools such as Google Docs \cite{google_google_2021-1} have implemented variations of these features, such as showing edit history and collaborators' cursor locations to increase workspace awareness. 
However, more research is warranted to explore how to integrate these features into future tools that support analysis of usability testing data like video recordings and meet the unique needs of UX practitioners.  

In terms of communication needs, more respondents preferred \textit{integrated voice/video call in analysis software} rather than \textit{integrated instant messaging}. 
However, current collaborative writing tools mainly employ the use of comments, while relying on external meeting software for synchronous collaboration. 
An explanation for the observed preference could be because voice/video calls are considered as \textit{richer} according to Media Richness Theory since it contains auditory (and possibly visual) signals that facilitate faster connections \cite{daft_organizational_1986}. 
In contrast, text-based media are considered less personal but more efficient when used during task-oriented conversations \cite{walther_computer-mediated_1996}. 
Another point to consider is that voice/video calls can only support synchronous collaboration whereas instant messaging supports both synchronous and asynchronous (where one can reply at a later time) \cite{mabrito_study_2006}. 
Future designers should consider these trade-offs when developing an integrated system for collaboration and data analysis as different forms of communication have been shown to affect patterns of interaction between collaborators and effectiveness of work outcome \cite{mabrito_study_2006, oviedo_meeting_2021}. 

\subsection{Implications of Prior UX Experience}

The correlations that we found in Sec. \ref{sec:correlations} suggest that prior UX experience is related to certain data analysis and collaboration practices. 
As years of experience grow, respondents were more likely to agree that \textit{it is time-consuming to conduct analysis on recordings} and also more likely to treat \textit{``improve reliability of results'' as a purpose for collaboration}. 
Higher awareness of the time-consuming nature of analysis and the importance of reliability could potentially lead to more thorough and robust analysis results. 
This is supported by prior studies showing that novice practitioners, who had not received training on user interface design principles, performed considerably worse than UX specialists and much worse than double specialists, i.e. those having expertise in user interface design and the domain of the software system \cite{nielsen_finding_1992}.
In relation to collaboration, survey respondents mentioned the lack of confidence in their colleagues' capabilities to analyze data as a key challenge.
Since the team consisted of varying levels of \textit{``maturity when it comes to UX understanding''}, it led to extra time needed to set the frame of analysis while collaborating. 

\subsubsection{Design Considerations Based on UX Experience}

\textbf{Effectively integrate team members with different qualifications and UX background.}
The behavioral model of group performance proposed by Sauer et al. states that the advantage of groups compared to individuals stems from increased task expertise \cite{sauer_effectiveness_2000}. 
Thus, the team performance may improve only if a team succeeds with integrating members with different qualifications, perspectives, and knowledge \cite{de_dreu_task_2003}. 
For example, teams could participate in certain interventions like \textit{team training}, which have been shown to improve work team effectiveness through goal-setting and team building \cite{buljac-samardzic_interventions_2010}. 
Future work, such as in-depth qualitative studies, can be conducted to further explore the reasons behind other collaboration trends observed with years of UX experience.

\rv{
\subsection{Impact of Emerging Technologies on UX Practices} 

\subsubsection{Recent Technological Changes}

Over the past ten years, technology has advanced at a rapid pace and UX practices must adapt to these fast-growing innovative environments.
There has been an emergence of start-up companies, which frequently guide their software development using agile practices \cite{paternoster_software_2014}. 
These companies are typically limited in human resources, leading to difficulties in hiring experienced UX practitioners and inadequate usability testing practices \cite{kuusinen_startup_2019}. 
One survey found that the top hindrance to UX practices in start-up companies was the short turnaround time for analysis \cite{silveira_ux_2021}, which is a challenge echoed by our survey results. Furthermore, 32\% of UX practitioners in start-up companies also experienced ``communication and collaboration gaps between UX and other professionals'' \cite{silveira_ux_2021}. Thus, how to make UX practices more lightweight to match agile practices and timeboxed sprints without compromising validity remains a key challenge \cite{hokkanen_minimum_2016, silveira_ux_2021,fan_practices_2020}.

In addition to changes in company structure, the popularization of everyday technology, such as wireless headphones, smartwatches, smart speakers, and VR headsets, has sparked new research on evaluating the UX of these products.
For example, researchers conducted focus groups and surveys to determine attributes affecting the UX of headphones \cite{jensen_analysis_2016},  longitudinal studies to improve smartwatch wearability \cite{jeong_smartwatch_2017}, and 
in-depth interviews to investigate the UX of smart speakers \cite{xiao_study_2018}. 
Furthermore, traditional questionnaires like the System Usability Scale (SUS) \cite{lewis_system_2018} were insufficient for VR headsets, which require additional considerations like motion sickness \cite{ames_development_2005}. Thus, researchers have developed custom scales for measuring the UX of VR headsets \cite{yu_evaluation_2019}. 
In term of analysis resources, our survey showed that the majority of the respondents still used Nielsen's usability heuristics \cite{nielsen_10_1994} and Norman's design principles \cite{norman_design_2002}. However, they also modified them to cater to their specific products. The need to update heuristics for new technology is evident in recent studies that utilized custom usability principles to evaluate smartwatches \cite{chun_qualitative_2018}. Recently, researchers also proposed a set of adapted heuristics for conversational agents \cite{langevin_heuristic_2021} and for VR products \cite{murtza_heuristic_2017}, based on testing and feedback by experts. 
Future work should develop sets of reusable heuristics that apply to new categories of products.

Aside from evaluating the UX of new products, recent advances in VR/AR technology have also been leveraged to develop new UX design and analysis methods \cite{kent_mixed_2021}. For example, respondents mentioned that they used physical post-its during analysis and found this feature missing in software tools. Researchers recently designed a VR system for writing and organizing virtual post-it notes that overcomes physical limitations like lack of space \cite{lee_post-post-it_2021}. Another AR application was developed to help users find specific post-its, cluster information, and visualize cluster metrics using AR overlays \cite{subramonyam_affinity_2019}. 
Although research demonstrated the effectiveness of such methods, they lack collaboration considerations and have limited adoption in industry. Thus, future work could explore how to enhance the collaboration capabilities of VR/AR systems for UX analysis and smoothly integrate them into existing practices. 

For the detection of UX problems, recent advances in AI show that it is possible to automate parts of this process. Researchers leveraged AI to assess the usability of digital interfaces  \cite{grigera_automatic_2017, oztekin_machine_2013-2, paterno_customizable_2017, harms_automated_2019, jeong_detecting_2020-1}. 
For example, user interaction events were utilized to create machine learning (ML) classifiers to detect usability issues of websites \cite{grigera_automatic_2017, paterno_customizable_2017}, mobile applications \cite{jeong_detecting_2020-1}, and VR applications \cite{harms_automated_2019}. 
However, these methods were primarily based on textual interaction logs, which are only available for specific types of interfaces. Depending on click and scroll maps limits their applicability to products without network connectivity such as coffee makers. 
In contrast, we found that respondents mainly used written notes, audio, and video recordings when conducting data analysis. 
Similar to how UX practitioners analyzed users' behaviors in recordings, researchers developed an AI that could infer when the participant encountered a problem during a think-aloud test session by analyzing the subtle speech patterns in their verbalizations \cite{fan_automatic_2020}.
While promising, such method needs improvements in accuracy and explainability to enhance users' trust \cite{liu_ai_2021}. Thus, to better help UX practitioners without disrupting their current workflow, AI systems might be designed to play an assistive role in analyzing the data and resources that UX practitioners are currently using including session recordings, written notes, and transcripts.

\subsubsection{Towards Human-AI Collaboration for UX Analysis}

Recent research has suggested that users exhibit subtle speech patterns such as abnormal pitch and speech rate when encountering usability problems \cite{fan_concurrent_2019}. 
Two-thirds of our respondents reviewed video recordings, which echoes the findings of prior work showing that UX practitioners use multi-modal information from both the acoustic and visual channels \cite{chilana_understanding_2010}.
When they attend to many signals simultaneously, they may not notice these subtle speech patterns. 
Thus, there is an opportunity for AI to assist practitioners in detecting and highlighting subtle speech patterns. 
In fact, recently, researchers began to design AI-assisted UX analysis tools, such as VisTA~\cite{fan_vista_2020} and CoUX \cite{soure_coux_2021}, to help UX practitioners with analyzing usability test sessions by presenting automatically-extracted problem-indicators to them.  
Although VisTA demonstrated the promise of human-AI collaboration for UX evaluation, it was designed to assist individual practitioners.
As a collaborative tool, CoUX was evaluated with pairs of UX practitioners, but the AI assistance was limited to video analysis \cite{soure_coux_2021}.
Thus, future work should investigate ways to integrate AI to support the human-human collaboration process for UX evaluation.
For example, one particular challenge experienced by over two-thirds of the respondents was merging analysis from multiple evaluators. 
The AI system could possibly act as a tie-breaker which is a strategy employed by almost one-quarter of the respondents. 
Another possible role for the AI system is to learn from conversations between practitioners and recognize disagreements to make appropriate suggestions. 
Thus, more research is warranted to determine whether and to what extent should AI be involved in the analysis process, since many human collaborators face challenges in understanding AI's capabilities or envisioning what it might be \cite{yang_investigating_2018}. 

In addition to the state-of-the-art research prototypes, some commercially available tools that our respondents used already contain features derived from AI and ML. For example,  analytical platforms like UserTesting \cite{usertesting_usertesting_2021} and PlaybookUX \cite{playbookux_playbookux_2021} offer sentiment analysis which detects whether the participant is expressing positive, negative, or neutral sentiment. Furthermore, UserTesting also provides ``smart tags'' where predefined situations such as confusion, dislike, or suggestion are automatically labelled \cite{usertesting_usertesting_2021}. 
There are also data-informed analytical tools like MixPanel \cite{mixpanel_2021} which provides correlation analysis on retention data, and UXTesting \cite{uxtesting_uxtesting_2022} which automatically detects participants' emotions.
Since these features are relatively new, their effectiveness is yet to be validated. Moreover, the UX community may benefit from integrating more subtle behavioral patterns (e.g., verbalization and speech indicators of UX problems~\cite{fan_concurrent_2019,Fan2021OlderAdults}) into existing tools. 
}

\section{Limitations and Future Work}
\label{sec:limitations}

Our survey respondents were from six continents with varying levels of UX experience. Thus, the findings are expected to provide a reasonably informative perspective of practices and challenges in collaboration during the data analysis phase among UX practitioners around the world. 
However, the number of responses from different continents was mismatched. While the majority of the data came from North America, Asia, and Europe, relatively fewer responses were from Latin America, Africa, and Oceania. Thus, one should be cautious about generalizing the findings to UX practitioners. Indeed, our research calls for more work to investigate UX practices and challenges in under-represented regions of our survey, such as countries in Africa, Latin America, and Oceania. \rv{Furthermore, studying and comparing the practices and challenges of UX practitioners in different countries could inform the UX community about the impact of culture on UX data analysis and collaboration, which will further help the community design customized collaborative UX data analysis strategies and tools. 

This survey was conducted during the COVID-19 pandemic, so there is a possibility that the findings would not be representative of all time. During the pandemic, certain issues may be at the forefront of the respondents' minds, such as remote collaboration being brought up as their greatest challenge. As articles have shown how the COVID-19 pandemic has changed product users \cite{moran_covid-19_2020}, future work can also explore its \mc{impacts} on UX practitioners. 
Furthermore, we focused on collaboration between UX practitioners within the same company and found correlations of analysis behaviors based on the in-house UX team size. However, cross-company collaboration is also a common practice for companies who cannot afford to hire expertise and maintain testing facilities \cite{douglas_global_2008} or have products that are sold in different countries or languages \cite{molich_tips_2004}. Prior research suggested that cross-company collaboration happened in the form of outsourcing participant recruitment \cite{molich_tips_2004}, small subtasks such as transcription \cite{bornoe_tagpad_2011}, or complete evaluations to usability testing services \cite{douglas_global_2008, molich_quest_2011}. These collaboration modes included tasks that were not directly related to data analysis, which was the focus of our research, but may lead to other challenges such as language barriers, sub-quality reports, and deviations from expected practices \cite{molich_quest_2011, molich_tips_2004, douglas_global_2008}. Thus, future work should further explore the state of cross-company collaborations in recent years with a focus on how the COVID-19 pandemic may have affected these practices.
}

In this survey, we focused on the analysis of data collected from usability testing, which only covers the pragmatic aspect of UX research. However, the UX community believes it is also valuable to consider the hedonic aspects, which refers to the momentary pleasures we experience when interacting with a product \cite{mekler_momentary_2016, van_schaik_modelling_2008}. Furthermore, there is research on how UX is related to eudaimonia, in which users gain meaning from need fulfillment \cite{mekler_momentary_2016} and how perceived beauty and aesthetics impact UX \cite{tuch_is_2012, van_schaik_modelling_2008}. Based on these aspects, future work should investigate how UX practitioners collect and analyze data for evaluating the overall product quality. From these findings, we can propose tools to better assist with discovering hedonic and eudaimonic issues in addition to task-oriented usability issues.

Our survey study primarily provides a quantitative understanding of UX practitioners' current practices, challenges, and opportunities in data analysis and collaboration even though it also offers qualitative insights via open-ended questions. To better understand the rich reasons behind the phenomena that our study revealed, more qualitative research, such as in-depth interviews with UX practitioners from different industrial sections in different continents, should be conducted. In particular, it is worth investigating different standards and principles that UX practitioners use for problem identification and merging and their considerations for choosing one over the other. 
Furthermore, as disagreements tend to occur during collaboration, it is valuable to gain a deeper understanding of their strategies for managing disagreements, in particular regarding whether there is a problem and what the severity level should be. 
Last but not least, as our study revealed that they would have a discussion with other practitioners to resolve disagreements, it would be interesting to explore the content of their conversations when discussing disagreements to better understand the nature of the disagreements and how they manage to gain a consensus. 
\section{Conclusion}
We have conducted an international survey to understand the practices and challenges of collaboration in the context of conducting data analysis on usability test sessions. Based on the responses of 279 participants who had varying UX experience and worked in different geographic locations, we found that UX practitioners collaborate to primarily \textit{identify more usability problems} and \textit{generate redesign suggestions}, and to \textit{improve reliability of results} to a lesser extent. 
We identified three modes of collaboration: \textit{independently analyze different portions of the data with little or no overlap and then collaborate (i.e., divide and conquer)}, \textit{collaboratively analyze the session with little or no independent analysis (i.e., group analysis)}, and \textit{independently analyze the same set of data and then collaborate}. Although the third mode was recommended by the literature, it was least adopted among our respondents because it was perceived to be more time-consuming than the other two. 
Moreover, most of the respondents encountered challenges related to lack of time as 70\% needed to complete analysis within 1 week, \mc{impacting their choice of collaboration modes}. 
These findings highlight an opportunity to address the trade-offs between efficiency and validity of analysis results. 
What's more, respondents also experienced disagreements with colleagues regarding usability problems and difficulty in merging analysis from multiple practitioners. 
Our survey findings could potentially inform UX practitioners about how their colleagues perceive collaboration during data analysis. 
In addition, our findings reveal opportunities for developing better methods and tools to facilitate collaboration \mc{during analysis}, for example, developing an integrated platform to support both analysis and collaboration in one place, establishing group analysis procedures, supporting both synchronous and asynchronous collaboration, and integrating team members with various skill levels.


\begin{acks}
We thank all our respondents for taking the time to share their experiences. 
\end{acks}

\bibliographystyle{ACM-Reference-Format}
\bibliography{bibs/main}


\begin{thebibliography}{116}


\ifx \showCODEN    \undefined \def \showCODEN     #1{\unskip}     \fi
\ifx \showDOI      \undefined \def \showDOI       #1{#1}\fi
\ifx \showISBNx    \undefined \def \showISBNx     #1{\unskip}     \fi
\ifx \showISBNxiii \undefined \def \showISBNxiii  #1{\unskip}     \fi
\ifx \showISSN     \undefined \def \showISSN      #1{\unskip}     \fi
\ifx \showLCCN     \undefined \def \showLCCN      #1{\unskip}     \fi
\ifx \shownote     \undefined \def \shownote      #1{#1}          \fi
\ifx \showarticletitle \undefined \def \showarticletitle #1{#1}   \fi
\ifx \showURL      \undefined \def \showURL       {\relax}        \fi
\providecommand\bibfield[2]{#2}
\providecommand\bibinfo[2]{#2}
\providecommand\natexlab[1]{#1}
\providecommand\showeprint[2][]{arXiv:#2}

\bibitem[\protect\citeauthoryear{Ames, Wolffsohn, and Mcbrien}{Ames
  et~al\mbox{.}}{2005}]%
        {ames_development_2005}
\bibfield{author}{\bibinfo{person}{Shelly~L. Ames}, \bibinfo{person}{James~S.
  Wolffsohn}, {and} \bibinfo{person}{Neville~A. Mcbrien}.}
  \bibinfo{year}{2005}\natexlab{}.
\newblock \showarticletitle{The Development of a Symptom Questionnaire for
  Assessing Virtual Reality Viewing Using a Head-Mounted Display}.
\newblock  \bibinfo{volume}{82}, \bibinfo{number}{3} (\bibinfo{year}{2005}),
  \bibinfo{pages}{168--176}.
\newblock
\showISSN{1538-9235}
\urldef\tempurl%
\url{https://doi.org/10.1097/01.OPX.0000156307.95086.6}
\showDOI{\tempurl}


\bibitem[\protect\citeauthoryear{Andre, Hartson, Belz, and McCreary}{Andre
  et~al\mbox{.}}{2001}]%
        {andre_user_2001}
\bibfield{author}{\bibinfo{person}{Terence Andre}, \bibinfo{person}{H Hartson},
  \bibinfo{person}{Steven Belz}, {and} \bibinfo{person}{Faith McCreary}.}
  \bibinfo{year}{2001}\natexlab{}.
\newblock \showarticletitle{User Action Framework: A Reliable Foundation for
  Usability Engineering Support Tools}.
\newblock \bibinfo{journal}{\emph{International Journal of Human-Computer
  Studies}}  \bibinfo{volume}{54} (\bibinfo{date}{Jan.} \bibinfo{year}{2001}),
  \bibinfo{pages}{107--136}.
\newblock
\urldef\tempurl%
\url{https://doi.org/10.1006/ijhc.2000.0441}
\showDOI{\tempurl}


\bibitem[\protect\citeauthoryear{Atlassian}{Atlassian}{2021a}]%
        {atlassian_confluence_2021}
\bibfield{author}{\bibinfo{person}{Atlassian}.}
  \bibinfo{year}{2021}\natexlab{a}.
\newblock \bibinfo{title}{Confluence: {{Your Remote}}-{{Friendly Team
  Workspace}}}.
\newblock
  \bibinfo{howpublished}{\url{https://www.atlassian.com/software/confluence}}.
\newblock


\bibitem[\protect\citeauthoryear{Atlassian}{Atlassian}{2021b}]%
        {atlassian_jira_2021}
\bibfield{author}{\bibinfo{person}{Atlassian}.}
  \bibinfo{year}{2021}\natexlab{b}.
\newblock \bibinfo{title}{Jira: {{Issue}} \& {{Project Tracking Software}}}.
\newblock
  \bibinfo{howpublished}{\url{https://www.atlassian.com/software/jira}}.
\newblock


\bibitem[\protect\citeauthoryear{Atlassian}{Atlassian}{2021c}]%
        {atlassian_trello_2021}
\bibfield{author}{\bibinfo{person}{Atlassian}.}
  \bibinfo{year}{2021}\natexlab{c}.
\newblock \bibinfo{title}{Trello}.
\newblock \bibinfo{howpublished}{\url{https://trello.com}}.
\newblock


\bibitem[\protect\citeauthoryear{Bertini, Gabrielli, Kimani, Catarci, and
  Santucci}{Bertini et~al\mbox{.}}{2006}]%
        {bertini_appropriating_2006-1}
\bibfield{author}{\bibinfo{person}{Enrico Bertini}, \bibinfo{person}{Silvia
  Gabrielli}, \bibinfo{person}{Stephen Kimani}, \bibinfo{person}{Tiziana
  Catarci}, {and} \bibinfo{person}{Giuseppe Santucci}.}
  \bibinfo{year}{2006}\natexlab{}.
\newblock \showarticletitle{Appropriating and Assessing Heuristics for Mobile
  Computing}. In \bibinfo{booktitle}{\emph{Proceedings of the Working
  Conference on {{Advanced}} Visual Interfaces}}
  \emph{(\bibinfo{series}{{{AVI}} '06})}. \bibinfo{publisher}{{Association for
  Computing Machinery}}, \bibinfo{address}{{New York, NY, USA}},
  \bibinfo{pages}{119--126}.
\newblock
\showISBNx{978-1-59593-353-9}
\urldef\tempurl%
\url{https://doi.org/10.1145/1133265.1133291}
\showDOI{\tempurl}


\bibitem[\protect\citeauthoryear{Boren and Ramey}{Boren and Ramey}{2000}]%
        {boren_thinking_2000}
\bibfield{author}{\bibinfo{person}{Ted Boren} {and} \bibinfo{person}{Judith
  Ramey}.} \bibinfo{year}{2000}\natexlab{}.
\newblock \showarticletitle{Thinking Aloud: Reconciling Theory and Practice}.
\newblock \bibinfo{journal}{\emph{IEEE Transactions on Professional
  Communication}} \bibinfo{volume}{43}, \bibinfo{number}{3}
  (\bibinfo{date}{Sept.} \bibinfo{year}{2000}), \bibinfo{pages}{261--278}.
\newblock
\showISSN{1558-1500}
\urldef\tempurl%
\url{https://doi.org/10.1109/47.867942}
\showDOI{\tempurl}


\bibitem[\protect\citeauthoryear{Bornoe, Barkhuus, Brown, and Hall}{Bornoe
  et~al\mbox{.}}{2011}]%
        {bornoe_tagpad_2011}
\bibfield{author}{\bibinfo{person}{Nis Bornoe}, \bibinfo{person}{Louise
  Barkhuus}, \bibinfo{person}{Barry Brown}, {and} \bibinfo{person}{Malcolm
  Hall}.} \bibinfo{year}{2011}\natexlab{}.
\newblock \showarticletitle{{TagPad} for {iPad} – Designing a Support Tool
  for Interview Studies}. In \bibinfo{booktitle}{\emph{11th Danish
  Human-Computer Interaction Research Symposium}} (Frederiksberg, Denmark).
  \bibinfo{pages}{49--52}.
\newblock
\showISBNx{978-87-92524-17-1}


\bibitem[\protect\citeauthoryear{Brynjolfsson, Horton, Ozimek, Rock, Sharma,
  and {TuYe}}{Brynjolfsson et~al\mbox{.}}{2020}]%
        {brynjolfsson_covid-19_2020}
\bibfield{author}{\bibinfo{person}{Erik Brynjolfsson}, \bibinfo{person}{John~J.
  Horton}, \bibinfo{person}{Adam Ozimek}, \bibinfo{person}{Daniel Rock},
  \bibinfo{person}{Garima Sharma}, {and} \bibinfo{person}{Hong-Yi {TuYe}}.}
  \bibinfo{year}{2020}\natexlab{}.
\newblock \bibinfo{title}{{COVID}-19 and Remote Work: An Early Look at {US}
  Data}.
\newblock
\newblock
\urldef\tempurl%
\url{https://doi.org/10.3386/w27344}
\showDOI{\tempurl}
\newblock
\shownote{Series: Working Paper Series.}


\bibitem[\protect\citeauthoryear{{Buljac-Samardzic}, {Dekker-van Doorn}, {van
  Wijngaarden}, and {van Wijk}}{{Buljac-Samardzic} et~al\mbox{.}}{2010}]%
        {buljac-samardzic_interventions_2010}
\bibfield{author}{\bibinfo{person}{Martina {Buljac-Samardzic}},
  \bibinfo{person}{Connie~M. {Dekker-van Doorn}}, \bibinfo{person}{Jeroen D.~H.
  {van Wijngaarden}}, {and} \bibinfo{person}{Kees~P. {van Wijk}}.}
  \bibinfo{year}{2010}\natexlab{}.
\newblock \showarticletitle{Interventions to Improve Team Effectiveness: {{A}}
  Systematic Review}.
\newblock \bibinfo{journal}{\emph{Health Policy}} \bibinfo{volume}{94},
  \bibinfo{number}{3} (\bibinfo{date}{March} \bibinfo{year}{2010}),
  \bibinfo{pages}{183--195}.
\newblock
\showISSN{0168-8510}
\urldef\tempurl%
\url{https://doi.org/10.1016/j.healthpol.2009.09.015}
\showDOI{\tempurl}


\bibitem[\protect\citeauthoryear{Capra}{Capra}{2006}]%
        {capra_usability_2006}
\bibfield{author}{\bibinfo{person}{Miranda~G. Capra}.}
  \bibinfo{year}{2006}\natexlab{}.
\newblock \emph{\bibinfo{title}{Usability {{Problem Description}} and the
  {{Evaluator Effect}} in {{Usability Testing}}}}.
\newblock \bibinfo{thesistype}{Ph.D. Dissertation}. \bibinfo{school}{Virginia
  Tech}, \bibinfo{address}{{Blacksburg, VA}}.
\newblock


\bibitem[\protect\citeauthoryear{Chilana, Wobbrock, and Ko}{Chilana
  et~al\mbox{.}}{2010}]%
        {chilana_understanding_2010}
\bibfield{author}{\bibinfo{person}{Parmit~K. Chilana},
  \bibinfo{person}{Jacob~O. Wobbrock}, {and} \bibinfo{person}{Andrew~J. Ko}.}
  \bibinfo{year}{2010}\natexlab{}.
\newblock \showarticletitle{Understanding Usability Practices in Complex
  Domains}. In \bibinfo{booktitle}{\emph{Proceedings of the 28th International
  Conference on {{Human}} Factors in Computing Systems - {{CHI}} '10}}.
  \bibinfo{publisher}{{ACM Press}}, \bibinfo{address}{{Atlanta, Georgia, USA}},
  \bibinfo{pages}{2337--2346}.
\newblock
\showISBNx{978-1-60558-929-9}
\urldef\tempurl%
\url{https://doi.org/10.1145/1753326.1753678}
\showDOI{\tempurl}


\bibitem[\protect\citeauthoryear{Chun, Dey, Lee, and Kim}{Chun
  et~al\mbox{.}}{2018}]%
        {chun_qualitative_2018}
\bibfield{author}{\bibinfo{person}{Jaemin Chun}, \bibinfo{person}{Anind Dey},
  \bibinfo{person}{Kyungtaek Lee}, {and} \bibinfo{person}{{SeungJun} Kim}.}
  \bibinfo{year}{2018}\natexlab{}.
\newblock \showarticletitle{A qualitative study of smartwatch usage and its
  usability}.
\newblock  \bibinfo{volume}{28}, \bibinfo{number}{4} (\bibinfo{year}{2018}),
  \bibinfo{pages}{186--199}.
\newblock
\showISSN{1520-6564}
\urldef\tempurl%
\url{https://doi.org/10.1002/hfm.20733}
\showDOI{\tempurl}
\newblock
\shownote{\_eprint: https://onlinelibrary.wiley.com/doi/pdf/10.1002/hfm.20733.}


\bibitem[\protect\citeauthoryear{Cockton, {Gilbert}, {Woolrych}, {Alan},
  {Hindmarch}, and {Mark}}{Cockton et~al\mbox{.}}{2004}]%
        {cockton_reconditioned_2004}
\bibfield{author}{\bibinfo{person}{Gilbert Cockton},
  \bibinfo{person}{{Gilbert}}, \bibinfo{person}{{Woolrych}},
  \bibinfo{person}{{Alan}}, \bibinfo{person}{{Hindmarch}}, {and}
  \bibinfo{person}{{Mark}}.} \bibinfo{year}{2004}\natexlab{}.
\newblock \showarticletitle{Reconditioned Merchandise: Extended Structured
  Report Formats in Usability Inspection}.
\newblock
\urldef\tempurl%
\url{https://doi.org/10.1145/985921.986083}
\showDOI{\tempurl}


\bibitem[\protect\citeauthoryear{Cockton and Lavery}{Cockton and
  Lavery}{1999}]%
        {cockton_framework_1999}
\bibfield{author}{\bibinfo{person}{Gilbert Cockton} {and}
  \bibinfo{person}{Darryn Lavery}.} \bibinfo{year}{1999}\natexlab{}.
\newblock \showarticletitle{A Framework for Usability Problem Extraction.}. In
  \bibinfo{booktitle}{\emph{{{INTERACT}}}}, Vol.~\bibinfo{volume}{1999}.
  \bibinfo{pages}{347--355}.
\newblock


\bibitem[\protect\citeauthoryear{Cohen}{Cohen}{1988}]%
        {cohen_statistical_1988}
\bibfield{author}{\bibinfo{person}{Jacob Cohen}.}
  \bibinfo{year}{1988}\natexlab{}.
\newblock \bibinfo{booktitle}{\emph{Statistical Power Analysis for the
  Behavioral Sciences} (\bibinfo{edition}{2nd ed} ed.)}.
\newblock \bibinfo{publisher}{{L. Erlbaum Associates}},
  \bibinfo{address}{{Hillsdale, N.J}}.
\newblock
\showISBNx{978-0-8058-0283-2}
\showLCCN{HA29 .C66 1988}


\bibitem[\protect\citeauthoryear{Daft and Lengel}{Daft and Lengel}{1986}]%
        {daft_organizational_1986}
\bibfield{author}{\bibinfo{person}{Richard Daft} {and} \bibinfo{person}{Robert
  Lengel}.} \bibinfo{year}{1986}\natexlab{}.
\newblock \showarticletitle{Organizational {{Information Requirements}},
  {{Media Richness}} and {{Structural Design}}}.
\newblock \bibinfo{journal}{\emph{Management Science}}  \bibinfo{volume}{32}
  (\bibinfo{date}{May} \bibinfo{year}{1986}), \bibinfo{pages}{554--571}.
\newblock
\urldef\tempurl%
\url{https://doi.org/10.1287/mnsc.32.5.554}
\showDOI{\tempurl}


\bibitem[\protect\citeauthoryear{De~Dreu and Weingart}{De~Dreu and
  Weingart}{2003}]%
        {de_dreu_task_2003}
\bibfield{author}{\bibinfo{person}{Carsten De~Dreu} {and}
  \bibinfo{person}{Laurie Weingart}.} \bibinfo{year}{2003}\natexlab{}.
\newblock \showarticletitle{Task {{Versus Relationship Conflict}}, {{Team
  Performance}}, and {{Team Member Satisfaction}}: {{A Meta}}-{{Analysis}}}.
\newblock \bibinfo{journal}{\emph{The Journal of applied psychology}}
  \bibinfo{volume}{88} (\bibinfo{date}{Sept.} \bibinfo{year}{2003}),
  \bibinfo{pages}{741--9}.
\newblock
\urldef\tempurl%
\url{https://doi.org/10.1037/0021-9010.88.4.741}
\showDOI{\tempurl}


\bibitem[\protect\citeauthoryear{de~Vreede, Fruhling, and Chakrapani}{de~Vreede
  et~al\mbox{.}}{2005}]%
        {de_vreede_repeatable_2005}
\bibfield{author}{\bibinfo{person}{Gert-Jan de Vreede}, \bibinfo{person}{Ann~L.
  Fruhling}, {and} \bibinfo{person}{Anita Chakrapani}.}
  \bibinfo{year}{2005}\natexlab{}.
\newblock \showarticletitle{A Repeatable Collaboration Process for Usability
  Testing}. In \bibinfo{booktitle}{\emph{Proceedings of the 38th Annual Hawaii
  International Conference on System Sciences}}. \bibinfo{pages}{46--46}.
\newblock
\urldef\tempurl%
\url{https://doi.org/10.1109/HICSS.2005.46}
\showDOI{\tempurl}
\newblock
\shownote{{ISSN}: 1530-1605.}


\bibitem[\protect\citeauthoryear{Dillman}{Dillman}{2000}]%
        {dillman_mail_2000}
\bibfield{author}{\bibinfo{person}{Don Dillman}.}
  \bibinfo{year}{2000}\natexlab{}.
\newblock \bibinfo{booktitle}{\emph{Mail and {{Internet Surveys}}: {{The
  Tailored Design Method}}}}. Vol.~\bibinfo{volume}{2}.
\newblock


\bibitem[\protect\citeauthoryear{Douglas}{Douglas}{2008}]%
        {douglas_global_2008}
\bibfield{author}{\bibinfo{person}{Ian Douglas}.}
  \bibinfo{year}{2008}\natexlab{}.
\newblock \showarticletitle{Global spread of usability expertise}. In
  \bibinfo{booktitle}{\emph{Proceedings of the 2008 Euro American Conference on
  Telematics and Information Systems}} (New York, {NY}, {USA})
  \emph{(\bibinfo{series}{{EATIS} '08})}. \bibinfo{publisher}{Association for
  Computing Machinery}, \bibinfo{pages}{1--4}.
\newblock
\showISBNx{978-1-59593-988-3}
\urldef\tempurl%
\url{https://doi.org/10.1145/1621087.1621115}
\showDOI{\tempurl}


\bibitem[\protect\citeauthoryear{Dumas and Loring}{Dumas and Loring}{2008}]%
        {dumas_moderating_2008}
\bibfield{author}{\bibinfo{person}{Joseph~S. Dumas} {and}
  \bibinfo{person}{Beth~A. Loring}.} \bibinfo{year}{2008}\natexlab{}.
\newblock \bibinfo{booktitle}{\emph{Moderating {{Usability Tests}}:
  {{Principles}} and {{Practices}} for {{Interacting}}}}.
\newblock \bibinfo{publisher}{{Elsevier}}.
\newblock
\showISBNx{978-0-08-055827-1}


\bibitem[\protect\citeauthoryear{Fan, Li, and Truong}{Fan
  et~al\mbox{.}}{2020a}]%
        {fan_automatic_2020}
\bibfield{author}{\bibinfo{person}{Mingming Fan}, \bibinfo{person}{Yue Li},
  {and} \bibinfo{person}{Khai~N. Truong}.} \bibinfo{year}{2020}\natexlab{a}.
\newblock \showarticletitle{Automatic {{Detection}} of {{Usability Problem
  Encounters}} in {{Think}}-Aloud {{Sessions}}}.
\newblock \bibinfo{journal}{\emph{ACM Transactions on Interactive Intelligent
  Systems}} \bibinfo{volume}{10}, \bibinfo{number}{2} (\bibinfo{date}{June}
  \bibinfo{year}{2020}), \bibinfo{pages}{1--24}.
\newblock
\showISSN{2160-6455, 2160-6463}
\urldef\tempurl%
\url{https://doi.org/10.1145/3385732}
\showDOI{\tempurl}


\bibitem[\protect\citeauthoryear{Fan, Lin, Chung, and Truong}{Fan
  et~al\mbox{.}}{2019}]%
        {fan_concurrent_2019}
\bibfield{author}{\bibinfo{person}{Mingming Fan}, \bibinfo{person}{Jinglan
  Lin}, \bibinfo{person}{Christina Chung}, {and} \bibinfo{person}{Khai~N.
  Truong}.} \bibinfo{year}{2019}\natexlab{}.
\newblock \showarticletitle{Concurrent {{Think}}-{{Aloud Verbalizations}} and
  {{Usability Problems}}}.
\newblock \bibinfo{journal}{\emph{ACM Transactions on Computer-Human
  Interaction}} \bibinfo{volume}{26}, \bibinfo{number}{5}
  (\bibinfo{date}{Sept.} \bibinfo{year}{2019}), \bibinfo{pages}{1--35}.
\newblock
\showISSN{1073-0516, 1557-7325}
\urldef\tempurl%
\url{https://doi.org/10.1145/3325281}
\showDOI{\tempurl}


\bibitem[\protect\citeauthoryear{Fan, Shi, and Truong}{Fan
  et~al\mbox{.}}{2020b}]%
        {fan_practices_2020}
\bibfield{author}{\bibinfo{person}{Mingming Fan}, \bibinfo{person}{Serina Shi},
  {and} \bibinfo{person}{Khai~N Truong}.} \bibinfo{year}{2020}\natexlab{b}.
\newblock \showarticletitle{Practices and {{Challenges}} of {{Using
  Think}}-{{Aloud Protocols}} in {{Industry}}: {{An International Survey}}}.
\newblock \bibinfo{journal}{\emph{Journal of Usability Studies}}
  \bibinfo{volume}{15}, \bibinfo{number}{2} (\bibinfo{year}{2020}),
  \bibinfo{pages}{85--102}.
\newblock


\bibitem[\protect\citeauthoryear{Fan, Wu, Zhao, Li, Wei, and Truong}{Fan
  et~al\mbox{.}}{2020c}]%
        {fan_vista_2020}
\bibfield{author}{\bibinfo{person}{Mingming Fan}, \bibinfo{person}{Ke Wu},
  \bibinfo{person}{Jian Zhao}, \bibinfo{person}{Yue Li},
  \bibinfo{person}{Winter Wei}, {and} \bibinfo{person}{Khai~N. Truong}.}
  \bibinfo{year}{2020}\natexlab{c}.
\newblock \showarticletitle{{{VisTA}}: {{Integrating Machine Intelligence}}
  with {{Visualization}} to {{Support}} the {{Investigation}} of
  {{Think}}-{{Aloud Sessions}}}.
\newblock \bibinfo{journal}{\emph{IEEE Transactions on Visualization and
  Computer Graphics}} \bibinfo{volume}{26}, \bibinfo{number}{1}
  (\bibinfo{date}{Jan.} \bibinfo{year}{2020}), \bibinfo{pages}{343--352}.
\newblock
\showISSN{1941-0506}
\urldef\tempurl%
\url{https://doi.org/10.1109/TVCG.2019.2934797}
\showDOI{\tempurl}


\bibitem[\protect\citeauthoryear{Fan, Zhao, and Tibdewal}{Fan
  et~al\mbox{.}}{2021}]%
        {Fan2021OlderAdults}
\bibfield{author}{\bibinfo{person}{Mingming Fan}, \bibinfo{person}{Qiwen Zhao},
  {and} \bibinfo{person}{Vinita Tibdewal}.} \bibinfo{year}{2021}\natexlab{}.
\newblock \showarticletitle{Older Adults’ Think-Aloud Verbalizations and
  Speech Features for Identifying User Experience Problems}. In
  \bibinfo{booktitle}{\emph{Proceedings of the 2021 CHI Conference on Human
  Factors in Computing Systems}} (Yokohama, Japan) \emph{(\bibinfo{series}{CHI
  '21})}. \bibinfo{publisher}{Association for Computing Machinery},
  \bibinfo{address}{New York, NY, USA}, Article \bibinfo{articleno}{358},
  \bibinfo{numpages}{13}~pages.
\newblock
\showISBNx{9781450380966}
\urldef\tempurl%
\url{https://doi.org/10.1145/3411764.3445680}
\showDOI{\tempurl}


\bibitem[\protect\citeauthoryear{Figma}{Figma}{2021}]%
        {figma_figma_2021}
\bibfield{author}{\bibinfo{person}{Figma}.} \bibinfo{year}{2021}\natexlab{}.
\newblock \bibinfo{title}{Figma: The Collaborative Interface Design Tool.}
\newblock \bibinfo{howpublished}{\url{https://www.figma.com/}}.
\newblock


\bibitem[\protect\citeauthoryear{F{\o}lstad, Law, and Hornb}{F{\o}lstad
  et~al\mbox{.}}{2012}]%
        {folstad_analysis_2012}
\bibfield{author}{\bibinfo{person}{Asbj{\o}rn F{\o}lstad},
  \bibinfo{person}{Effie Lai-Chong Law}, {and} \bibinfo{person}{Kasper Hornb}.}
  \bibinfo{year}{2012}\natexlab{}.
\newblock \showarticletitle{Analysis in Practical Usability Evaluation: A
  Survey Study}. In \bibinfo{booktitle}{\emph{Proceedings of the 30th {{SIGCHI
  Conference}} on {{Human Factors}} in {{Computing Systems}} - {{CHI}} '12}}.
  \bibinfo{publisher}{{ACM Press}}, \bibinfo{address}{{Austin, Texas}},
  \bibinfo{pages}{2127--2136}.
\newblock
\urldef\tempurl%
\url{https://doi.org/10.1145/2207676.2208365}
\showDOI{\tempurl}


\bibitem[\protect\citeauthoryear{F{\o}lstad, Law, and Hornb{\ae}k}{F{\o}lstad
  et~al\mbox{.}}{2010}]%
        {folstad_analysis_2010}
\bibfield{author}{\bibinfo{person}{Asbj{\o}rn F{\o}lstad},
  \bibinfo{person}{Effie Lai-Chong Law}, {and} \bibinfo{person}{Kasper
  Hornb{\ae}k}.} \bibinfo{year}{2010}\natexlab{}.
\newblock \showarticletitle{Analysis in Usability Evaluations: An Exploratory
  Study}. In \bibinfo{booktitle}{\emph{Proceedings of the 6th {{Nordic
  Conference}} on {{Human}}-{{Computer Interaction}}: {{Extending
  Boundaries}}}} \emph{(\bibinfo{series}{{{NordiCHI}} '10})}.
  \bibinfo{publisher}{{Association for Computing Machinery}},
  \bibinfo{address}{{New York, NY, USA}}, \bibinfo{pages}{647--650}.
\newblock
\showISBNx{978-1-60558-934-3}
\urldef\tempurl%
\url{https://doi.org/10.1145/1868914.1868995}
\showDOI{\tempurl}


\bibitem[\protect\citeauthoryear{Fruhling and de~Vreede}{Fruhling and
  de~Vreede}{2006}]%
        {fruhling_collaborative_2006}
\bibfield{author}{\bibinfo{person}{Ann~L. Fruhling} {and}
  \bibinfo{person}{Gert-Jan de Vreede}.} \bibinfo{year}{2006}\natexlab{}.
\newblock \showarticletitle{Collaborative Usability Testing to Facilitate
  Stakeholder Involvement}.
\newblock In \bibinfo{booktitle}{\emph{Value-Based Software Engineering}},
  \bibfield{editor}{\bibinfo{person}{Stefan Biffl}, \bibinfo{person}{Aybüke
  Aurum}, \bibinfo{person}{Barry Boehm}, \bibinfo{person}{Hakan Erdogmus},
  {and} \bibinfo{person}{Paul Grünbacher}} (Eds.).
  \bibinfo{publisher}{Springer}.
\newblock
\showISBNx{978-3-540-29263-0}
\urldef\tempurl%
\url{https://doi.org/10.1007/3-540-29263-2_10}
\showDOI{\tempurl}


\bibitem[\protect\citeauthoryear{FullStory}{FullStory}{2021}]%
        {fullstory_fullstory_2021}
\bibfield{author}{\bibinfo{person}{FullStory}.}
  \bibinfo{year}{2021}\natexlab{}.
\newblock \bibinfo{title}{{{FullStory}} | {{Robust Analytics}}, {{Session
  Replay}}, {{Heatmaps}}, {{Dev Tools}}}.
\newblock \bibinfo{howpublished}{\url{https://www.fullstory.com}}.
\newblock


\bibitem[\protect\citeauthoryear{Google}{Google}{2021}]%
        {google_google_2021-1}
\bibfield{author}{\bibinfo{person}{Google}.} \bibinfo{year}{2021}\natexlab{}.
\newblock \bibinfo{title}{Google {{Workspace}}: {{Business Collaboration
  Tools}}}.
\newblock \bibinfo{howpublished}{\url{https://workspace.google.com}}.
\newblock


\bibitem[\protect\citeauthoryear{Grigera, Garrido, Rivero, and Rossi}{Grigera
  et~al\mbox{.}}{2017}]%
        {grigera_automatic_2017}
\bibfield{author}{\bibinfo{person}{J. Grigera}, \bibinfo{person}{Alejandra
  Garrido}, \bibinfo{person}{J. Rivero}, {and} \bibinfo{person}{G. Rossi}.}
  \bibinfo{year}{2017}\natexlab{}.
\newblock \showarticletitle{Automatic Detection of Usability Smells in Web
  Applications}.
\newblock \bibinfo{journal}{\emph{Int. J. Hum. Comput. Stud.}}
  (\bibinfo{year}{2017}).
\newblock
\urldef\tempurl%
\url{https://doi.org/10.1016/j.ijhcs.2016.09.009}
\showDOI{\tempurl}


\bibitem[\protect\citeauthoryear{Gutwin and Greenberg}{Gutwin and
  Greenberg}{2002}]%
        {gutwin_descriptive_2002}
\bibfield{author}{\bibinfo{person}{Carl Gutwin} {and} \bibinfo{person}{Saul
  Greenberg}.} \bibinfo{year}{2002}\natexlab{}.
\newblock \showarticletitle{A {{Descriptive Framework}} of {{Workspace
  Awareness}} for {{Real}}-{{Time Groupware}}}.
\newblock \bibinfo{journal}{\emph{Computer Supported Cooperative Work}}
  \bibinfo{volume}{11}, \bibinfo{number}{3} (\bibinfo{date}{Nov.}
  \bibinfo{year}{2002}), \bibinfo{pages}{411--446}.
\newblock
\showISSN{0925-9724}
\urldef\tempurl%
\url{https://doi.org/10.1023/A:1021271517844}
\showDOI{\tempurl}


\bibitem[\protect\citeauthoryear{Harms}{Harms}{2019}]%
        {harms_automated_2019}
\bibfield{author}{\bibinfo{person}{Patrick Harms}.}
  \bibinfo{year}{2019}\natexlab{}.
\newblock \showarticletitle{Automated {{Usability Evaluation}} of {{Virtual
  Reality Applications}}}.
\newblock \bibinfo{journal}{\emph{ACM Transactions on Computer-Human
  Interaction}} \bibinfo{volume}{26}, \bibinfo{number}{3}
  (\bibinfo{date}{April} \bibinfo{year}{2019}), \bibinfo{pages}{14:1--14:36}.
\newblock
\showISSN{1073-0516}
\urldef\tempurl%
\url{https://doi.org/10.1145/3301423}
\showDOI{\tempurl}


\bibitem[\protect\citeauthoryear{Hartson, Andre, and Williges}{Hartson
  et~al\mbox{.}}{2003}]%
        {hartson_criteria_2003}
\bibfield{author}{\bibinfo{person}{H.~Rex Hartson}, \bibinfo{person}{Terence
  Andre}, {and} \bibinfo{person}{Robert Williges}.}
  \bibinfo{year}{2003}\natexlab{}.
\newblock \showarticletitle{Criteria {{For Evaluating Usability Evaluation
  Methods}}}.
\newblock \bibinfo{journal}{\emph{Int. J. Hum. Comput. Interaction}}
  \bibinfo{volume}{15} (\bibinfo{date}{Feb.} \bibinfo{year}{2003}),
  \bibinfo{pages}{145--181}.
\newblock
\urldef\tempurl%
\url{https://doi.org/10.1207/S15327590IJHC1501_13}
\showDOI{\tempurl}


\bibitem[\protect\citeauthoryear{Hertzum and Jacobsen}{Hertzum and
  Jacobsen}{2001}]%
        {hertzum_evaluator_2001}
\bibfield{author}{\bibinfo{person}{Morten Hertzum} {and}
  \bibinfo{person}{Niels~Ebbe Jacobsen}.} \bibinfo{year}{2001}\natexlab{}.
\newblock \showarticletitle{The {{Evaluator Effect}}: {{A Chilling Fact About
  Usability Evaluation Methods}}}.
\newblock \bibinfo{journal}{\emph{International Journal of Human-Computer
  Interaction}} \bibinfo{volume}{15}, \bibinfo{number}{1}
  (\bibinfo{year}{2001}), \bibinfo{pages}{183--204}.
\newblock
\showISSN{1044-7318, 1532-7590}
\urldef\tempurl%
\url{https://doi.org/10.1207/S15327590IJHC1501_14}
\showDOI{\tempurl}


\bibitem[\protect\citeauthoryear{Hertzum, Molich, and Jacobsen}{Hertzum
  et~al\mbox{.}}{2013}]%
        {hertzum_what_2013}
\bibfield{author}{\bibinfo{person}{Morten Hertzum}, \bibinfo{person}{Rolf
  Molich}, {and} \bibinfo{person}{Niels~Ebbe Jacobsen}.}
  \bibinfo{year}{2013}\natexlab{}.
\newblock \showarticletitle{What You Get Is What You See: Revisiting the
  Evaluator Effect in Usability Tests}.
\newblock \bibinfo{journal}{\emph{Behaviour \& Information Technology}}
  \bibinfo{volume}{33}, \bibinfo{number}{2} (\bibinfo{date}{April}
  \bibinfo{year}{2013}), \bibinfo{pages}{144--162}.
\newblock
\showISSN{0144-929X}
\urldef\tempurl%
\url{https://doi.org/10.1080/0144929X.2013.783114}
\showDOI{\tempurl}


\bibitem[\protect\citeauthoryear{Hoffmann, J{\'o}nsd{\'o}ttir, and
  Hvannberg}{Hoffmann et~al\mbox{.}}{2019}]%
        {hoffmann_consolidation_2019}
\bibfield{author}{\bibinfo{person}{Rebekka Hoffmann},
  \bibinfo{person}{Anna~Helga J{\'o}nsd{\'o}ttir}, {and}
  \bibinfo{person}{Ebba~Thora Hvannberg}.} \bibinfo{year}{2019}\natexlab{}.
\newblock \showarticletitle{Consolidation of {{Usability Problems With Novice
  Evaluators Re}}-{{Examined}} in {{Individual}} vs. {{Collaborative
  Settings}}}.
\newblock \bibinfo{journal}{\emph{Interacting with Computers}}
  \bibinfo{volume}{31}, \bibinfo{number}{6} (\bibinfo{date}{April}
  \bibinfo{year}{2019}), \bibinfo{pages}{525--538}.
\newblock
\showISSN{0953-5438}
\urldef\tempurl%
\url{https://doi.org/10.1093/iwc/iwz034}
\showDOI{\tempurl}


\bibitem[\protect\citeauthoryear{Hokkanen, Kuusinen, and Väänänen}{Hokkanen
  et~al\mbox{.}}{2016}]%
        {hokkanen_minimum_2016}
\bibfield{author}{\bibinfo{person}{Laura Hokkanen}, \bibinfo{person}{Kati
  Kuusinen}, {and} \bibinfo{person}{Kaisa Väänänen}.}
  \bibinfo{year}{2016}\natexlab{}.
\newblock \showarticletitle{Minimum Viable User {EXperience}: A Framework for
  Supporting Product Design in Startups}. In \bibinfo{booktitle}{\emph{Agile
  Processes, in Software Engineering, and Extreme Programming}} (Cham)
  \emph{(\bibinfo{series}{Lecture Notes in Business Information Processing})},
  \bibfield{editor}{\bibinfo{person}{Helen Sharp} {and} \bibinfo{person}{Tracy
  Hall}} (Eds.). \bibinfo{publisher}{Springer International Publishing},
  \bibinfo{pages}{66--78}.
\newblock
\showISBNx{978-3-319-33515-5}
\urldef\tempurl%
\url{https://doi.org/10.1007/978-3-319-33515-5_6}
\showDOI{\tempurl}


\bibitem[\protect\citeauthoryear{Inc.}{Inc.}{2020}]%
        {tactivos_inc_mural_2020}
\bibfield{author}{\bibinfo{person}{Tactivos Inc.}}
  \bibinfo{year}{2020}\natexlab{}.
\newblock \bibinfo{title}{{{MURAL}}: A Digital Workspace for Visual
  Collaboration}.
\newblock \bibinfo{howpublished}{\url{https://www.mural.co/}}.
\newblock


\bibitem[\protect\citeauthoryear{{ISO}}{{ISO}}{2020}]%
        {iso_iso_2020}
\bibfield{author}{\bibinfo{person}{{ISO}}.} \bibinfo{year}{2020}\natexlab{}.
\newblock \bibinfo{title}{{{ISO}} 9241-110:2020 {{Ergonomics}} of Human-System
  Interaction}.
\newblock
  \bibinfo{howpublished}{\url{https://www.iso.org/cms/render/live/en/sites/isoorg/contents/data/standard/07/52/75258.html}}.
\newblock


\bibitem[\protect\citeauthoryear{Jacobsen, E, Hertzum, {Morten}, John, and
  E}{Jacobsen et~al\mbox{.}}{1998}]%
        {jacobsen_evaluator_1998}
\bibfield{author}{\bibinfo{person}{Niels Jacobsen}, \bibinfo{person}{Niels E},
  \bibinfo{person}{Morten Hertzum}, \bibinfo{person}{{Morten}},
  \bibinfo{person}{Bonnie John}, {and} \bibinfo{person}{Bonnie E}.}
  \bibinfo{year}{1998}\natexlab{}.
\newblock \showarticletitle{The Evaluator Effect in Usability Tests}.
\newblock
\urldef\tempurl%
\url{https://doi.org/10.1145/286498.286737}
\showDOI{\tempurl}


\bibitem[\protect\citeauthoryear{Jensen, Lauridsen, Poulsen, Tofte, and
  Christensen}{Jensen et~al\mbox{.}}{2016}]%
        {jensen_analysis_2016}
\bibfield{author}{\bibinfo{person}{Rasmus Jensen}, \bibinfo{person}{Nikolaj
  Lauridsen}, \bibinfo{person}{Andreas Poulsen}, \bibinfo{person}{Casper
  Tofte}, {and} \bibinfo{person}{Flemming Christensen}.}
  \bibinfo{year}{2016}\natexlab{}.
\newblock \showarticletitle{Analysis of Subjective Evaluation of User
  Experience with Headphones}. \bibinfo{publisher}{Audio Engineering Society}.
\newblock
\urldef\tempurl%
\url{https://www.aes.org/e-lib/browse.cfm?elib=18345}
\showURL{%
\tempurl}


\bibitem[\protect\citeauthoryear{Jeong, Kim, Kim, Lee, and Jeong}{Jeong
  et~al\mbox{.}}{2017}]%
        {jeong_smartwatch_2017}
\bibfield{author}{\bibinfo{person}{Hayeon Jeong}, \bibinfo{person}{Heepyung
  Kim}, \bibinfo{person}{Rihun Kim}, \bibinfo{person}{Uichin Lee}, {and}
  \bibinfo{person}{Yong Jeong}.} \bibinfo{year}{2017}\natexlab{}.
\newblock \showarticletitle{Smartwatch Wearing Behavior Analysis: A
  Longitudinal Study}.
\newblock  \bibinfo{volume}{1}, \bibinfo{number}{3} (\bibinfo{year}{2017}),
  \bibinfo{pages}{60:1--60:31}.
\newblock
\urldef\tempurl%
\url{https://doi.org/10.1145/3131892}
\showDOI{\tempurl}


\bibitem[\protect\citeauthoryear{Jeong, Kim, and In}{Jeong
  et~al\mbox{.}}{2020}]%
        {jeong_detecting_2020-1}
\bibfield{author}{\bibinfo{person}{JongWook Jeong}, \bibinfo{person}{NeungHoe
  Kim}, {and} \bibinfo{person}{Hoh~Peter In}.} \bibinfo{year}{2020}\natexlab{}.
\newblock \showarticletitle{Detecting Usability Problems in Mobile Applications
  on the Basis of Dissimilarity in User Behavior}.
\newblock \bibinfo{journal}{\emph{International Journal of Human-Computer
  Studies}}  \bibinfo{volume}{139} (\bibinfo{date}{July} \bibinfo{year}{2020}),
  \bibinfo{pages}{102364}.
\newblock
\showISSN{1071-5819}
\urldef\tempurl%
\url{https://doi.org/10.1016/j.ijhcs.2019.10.001}
\showDOI{\tempurl}


\bibitem[\protect\citeauthoryear{Jones and Thoma}{Jones and Thoma}{2019}]%
        {jones_determinants_2019}
\bibfield{author}{\bibinfo{person}{Alexander Jones} {and}
  \bibinfo{person}{Volker Thoma}.} \bibinfo{year}{2019}\natexlab{}.
\newblock \showarticletitle{Determinants for {{Successful Agile Collaboration}}
  between {{UX Designers}} and {{Software Developers}} in a {{Complex
  Organisation}}}.
\newblock \bibinfo{journal}{\emph{International Journal of Human\textendash
  Computer Interaction}} \bibinfo{volume}{35}, \bibinfo{number}{20}
  (\bibinfo{date}{Dec.} \bibinfo{year}{2019}), \bibinfo{pages}{1914--1935}.
\newblock
\showISSN{1044-7318}
\urldef\tempurl%
\url{https://doi.org/10.1080/10447318.2019.1587856}
\showDOI{\tempurl}


\bibitem[\protect\citeauthoryear{Kent, Snider, Gopsill, and Hicks}{Kent
  et~al\mbox{.}}{2021}]%
        {kent_mixed_2021}
\bibfield{author}{\bibinfo{person}{Lee Kent}, \bibinfo{person}{Chris Snider},
  \bibinfo{person}{James Gopsill}, {and} \bibinfo{person}{Ben Hicks}.}
  \bibinfo{year}{2021}\natexlab{}.
\newblock \showarticletitle{Mixed reality in design prototyping: A systematic
  review}.
\newblock   \bibinfo{volume}{77} (\bibinfo{year}{2021}).
\newblock
\urldef\tempurl%
\url{https://doi.org/10.1016/j.destud.2021.101046}
\showDOI{\tempurl}


\bibitem[\protect\citeauthoryear{Kjeldskov, {Jesper}, Skov, B, {Stage}, and
  {Jan}}{Kjeldskov et~al\mbox{.}}{2004}]%
        {kjeldskov_instant_2004}
\bibfield{author}{\bibinfo{person}{Jesper Kjeldskov},
  \bibinfo{person}{{Jesper}}, \bibinfo{person}{Mikael Skov},
  \bibinfo{person}{Mikael B}, \bibinfo{person}{{Stage}}, {and}
  \bibinfo{person}{{Jan}}.} \bibinfo{year}{2004}\natexlab{}.
\newblock \showarticletitle{Instant Data Analysis: Conducting Usability
  Evaluations in a Day}.
\newblock
\urldef\tempurl%
\url{https://doi.org/10.1145/1028014.1028050}
\showDOI{\tempurl}


\bibitem[\protect\citeauthoryear{Koffka}{Koffka}{2013}]%
        {koffka_principles_2013}
\bibfield{author}{\bibinfo{person}{Kurt Koffka}.}
  \bibinfo{year}{2013}\natexlab{}.
\newblock \bibinfo{booktitle}{\emph{Principles {{Of Gestalt Psychology}}}}.
\newblock \bibinfo{publisher}{{Routledge}}.
\newblock
\showISBNx{978-1-136-30681-5}


\bibitem[\protect\citeauthoryear{Kuusinen, Sørensen, Frederiksen, Laugesen,
  and Juul}{Kuusinen et~al\mbox{.}}{2019}]%
        {kuusinen_startup_2019}
\bibfield{author}{\bibinfo{person}{Kati Kuusinen},
  \bibinfo{person}{Martin~Kjølbye Sørensen}, \bibinfo{person}{Nicklas~Mandrup
  Frederiksen}, \bibinfo{person}{Niclas~Kildahl Laugesen}, {and}
  \bibinfo{person}{Søren~Holm Juul}.} \bibinfo{year}{2019}\natexlab{}.
\newblock \showarticletitle{From Startup to Scaleup: An Interview Study of the
  Development of User Experience Work in a Data-Intensive Company}. In
  \bibinfo{booktitle}{\emph{Human-Centered Software Engineering}} (Cham)
  \emph{(\bibinfo{series}{Lecture Notes in Computer Science})},
  \bibfield{editor}{\bibinfo{person}{Cristian Bogdan}, \bibinfo{person}{Kati
  Kuusinen}, \bibinfo{person}{Marta~Kristín Lárusdóttir},
  \bibinfo{person}{Philippe Palanque}, {and} \bibinfo{person}{Marco Winckler}}
  (Eds.). \bibinfo{publisher}{Springer International Publishing},
  \bibinfo{pages}{3--14}.
\newblock
\showISBNx{978-3-030-05909-5}
\urldef\tempurl%
\url{https://doi.org/10.1007/978-3-030-05909-5_1}
\showDOI{\tempurl}


\bibitem[\protect\citeauthoryear{Langevin, Lordon, Avrahami, Cowan, Hirsch, and
  Hsieh}{Langevin et~al\mbox{.}}{2021}]%
        {langevin_heuristic_2021}
\bibfield{author}{\bibinfo{person}{Raina Langevin}, \bibinfo{person}{Ross~J
  Lordon}, \bibinfo{person}{Thi Avrahami}, \bibinfo{person}{Benjamin~R. Cowan},
  \bibinfo{person}{Tad Hirsch}, {and} \bibinfo{person}{Gary Hsieh}.}
  \bibinfo{year}{2021}\natexlab{}.
\newblock \showarticletitle{Heuristic Evaluation of Conversational Agents}.
\newblock  \bibinfo{number}{632} (\bibinfo{year}{2021}),
  \bibinfo{pages}{1--15}.
\newblock
\showISBNx{978-1-4503-8096-6}
\urldef\tempurl%
\url{https://doi.org/10.1145/3411764.3445312}
\showURL{%
\tempurl}


\bibitem[\protect\citeauthoryear{Lavery, Cockton, and Atkinson}{Lavery
  et~al\mbox{.}}{1997}]%
        {lavery_comparison_1997}
\bibfield{author}{\bibinfo{person}{Darryn Lavery}, \bibinfo{person}{Gilbert
  Cockton}, {and} \bibinfo{person}{Malcolm~P. Atkinson}.}
  \bibinfo{year}{1997}\natexlab{}.
\newblock \showarticletitle{Comparison of Evaluation Methods Using Structured
  Usability Problem Reports}.
\newblock \bibinfo{journal}{\emph{Behaviour \& Information Technology}}
  \bibinfo{volume}{16}, \bibinfo{number}{4-5} (\bibinfo{date}{Jan.}
  \bibinfo{year}{1997}), \bibinfo{pages}{246--266}.
\newblock
\showISSN{0144-929X}
\urldef\tempurl%
\url{https://doi.org/10.1080/014492997119824}
\showDOI{\tempurl}


\bibitem[\protect\citeauthoryear{Law and Hvannberg}{Law and Hvannberg}{2004}]%
        {law_analysis_2004}
\bibfield{author}{\bibinfo{person}{Effie Lai-Chong Law} {and}
  \bibinfo{person}{Ebba~Thora Hvannberg}.} \bibinfo{year}{2004}\natexlab{}.
\newblock \showarticletitle{Analysis of Strategies for Improving and Estimating
  the Effectiveness of Heuristic Evaluation}. In
  \bibinfo{booktitle}{\emph{Proceedings of the Third {{Nordic}} Conference on
  {{Human}}-Computer Interaction}} \emph{(\bibinfo{series}{{{NordiCHI}} '04})}.
  \bibinfo{publisher}{{Association for Computing Machinery}},
  \bibinfo{address}{{New York, NY, USA}}, \bibinfo{pages}{241--250}.
\newblock
\showISBNx{978-1-58113-857-3}
\urldef\tempurl%
\url{https://doi.org/10.1145/1028014.1028051}
\showDOI{\tempurl}


\bibitem[\protect\citeauthoryear{Law and Hvannberg}{Law and Hvannberg}{2008}]%
        {law_consolidating_2008}
\bibfield{author}{\bibinfo{person}{Effie Lai-Chong Law} {and}
  \bibinfo{person}{Ebba~Thora Hvannberg}.} \bibinfo{year}{2008}\natexlab{}.
\newblock \showarticletitle{Consolidating Usability Problems with Novice
  Evaluators}. In \bibinfo{booktitle}{\emph{Proceedings of the 5th {{Nordic}}
  Conference on {{Human}}-Computer Interaction: Building Bridges}}
  \emph{(\bibinfo{series}{{{NordiCHI}} '08})}. \bibinfo{publisher}{{Association
  for Computing Machinery}}, \bibinfo{address}{{New York, NY, USA}},
  \bibinfo{pages}{495--498}.
\newblock
\showISBNx{978-1-59593-704-9}
\urldef\tempurl%
\url{https://doi.org/10.1145/1463160.1463228}
\showDOI{\tempurl}


\bibitem[\protect\citeauthoryear{Lee, Ma, Cho, and Bae}{Lee
  et~al\mbox{.}}{2021}]%
        {lee_post-post-it_2021}
\bibfield{author}{\bibinfo{person}{Joon~Hyub Lee}, \bibinfo{person}{Donghyeok
  Ma}, \bibinfo{person}{Haena Cho}, {and} \bibinfo{person}{Seok-Hyung Bae}.}
  \bibinfo{year}{2021}\natexlab{}.
\newblock \showarticletitle{Post-Post-it: A Spatial Ideation System in {VR} for
  Overcoming Limitations of Physical Post-it Notes}.
\newblock In \bibinfo{booktitle}{\emph{Extended Abstracts of the 2021 {CHI}
  Conference on Human Factors in Computing Systems}}. Number 300.
  \bibinfo{publisher}{Association for Computing Machinery},
  \bibinfo{pages}{1--7}.
\newblock
\showISBNx{978-1-4503-8095-9}
\urldef\tempurl%
\url{https://doi.org/10.1145/3411763.3451786}
\showURL{%
\tempurl}


\bibitem[\protect\citeauthoryear{Lewis}{Lewis}{2006}]%
        {lewis_usability_2006}
\bibfield{author}{\bibinfo{person}{James~R. Lewis}.}
  \bibinfo{year}{2006}\natexlab{}.
\newblock \showarticletitle{Usability {{Testing}}}.
\newblock In \bibinfo{booktitle}{\emph{Handbook of {{Human Factors}} and
  {{Ergonomics}}}}. \bibinfo{publisher}{{John Wiley \& Sons, Ltd}}, Chapter~46,
  \bibinfo{pages}{1267--1312}.
\newblock
\showISBNx{978-1-118-13135-0}
\urldef\tempurl%
\url{https://doi.org/10.1002/9781118131350.ch46}
\showDOI{\tempurl}


\bibitem[\protect\citeauthoryear{Lewis}{Lewis}{2018}]%
        {lewis_system_2018}
\bibfield{author}{\bibinfo{person}{James~R. Lewis}.}
  \bibinfo{year}{2018}\natexlab{}.
\newblock \showarticletitle{The System Usability Scale: Past, Present, and
  Future}.
\newblock  \bibinfo{volume}{34}, \bibinfo{number}{7} (\bibinfo{year}{2018}),
  \bibinfo{pages}{577--590}.
\newblock
\showISSN{1044-7318}
\urldef\tempurl%
\url{https://doi.org/10.1080/10447318.2018.1455307}
\showDOI{\tempurl}
\newblock
\shownote{Publisher: Taylor \& Francis.}


\bibitem[\protect\citeauthoryear{Liu}{Liu}{2021}]%
        {liu_ai_2021}
\bibfield{author}{\bibinfo{person}{Bingjie Liu}.}
  \bibinfo{year}{2021}\natexlab{}.
\newblock \showarticletitle{In {AI} We Trust? Effects of Agency Locus and
  Transparency on Uncertainty Reduction in Human–{AI} Interaction}.
\newblock  (\bibinfo{year}{2021}).
\newblock
Issue zmab013.
\showISSN{1083-6101}
\urldef\tempurl%
\url{https://doi.org/10.1093/jcmc/zmab013}
\showDOI{\tempurl}


\bibitem[\protect\citeauthoryear{Lowry, Albrecht, Lee, and Nunamaker}{Lowry
  et~al\mbox{.}}{2002}]%
        {lowry_users_2002}
\bibfield{author}{\bibinfo{person}{Paul~B. Lowry}, \bibinfo{person}{Conan
  Albrecht}, \bibinfo{person}{James~D. Lee}, {and} \bibinfo{person}{Jay~F.
  Nunamaker}.} \bibinfo{year}{2002}\natexlab{}.
\newblock \showarticletitle{Users Experiences in Collaborative Writing Using
  Collaboratus: An Internet-Based Collaborative Work}.
  \bibinfo{publisher}{{IEEE} Computer Society}, \bibinfo{pages}{21--21}.
\newblock
\showISBNx{978-0-7695-1435-2}
\urldef\tempurl%
\url{https://doi.org/10.1109/HICSS.2002.993879}
\showDOI{\tempurl}


\bibitem[\protect\citeauthoryear{Lowry and Roberts}{Lowry and Roberts}{2003}]%
        {lowry_improving_2003}
\bibfield{author}{\bibinfo{person}{Paul~B. Lowry} {and} \bibinfo{person}{Tom~L.
  Roberts}.} \bibinfo{year}{2003}\natexlab{}.
\newblock \showarticletitle{Improving the Usability Evaluation Technique,
  Heuristic Evaluation, Through the Use of Collaborative Software}.
  \bibinfo{pages}{284}.
\newblock
\urldef\tempurl%
\url{https://doi.org/10.2139/ssrn.666224}
\showDOI{\tempurl}


\bibitem[\protect\citeauthoryear{Mabrito}{Mabrito}{2006}]%
        {mabrito_study_2006}
\bibfield{author}{\bibinfo{person}{Mark Mabrito}.}
  \bibinfo{year}{2006}\natexlab{}.
\newblock \showarticletitle{A {{Study}} of {{Synchronous Versus Asynchronous
  Collaboration}} in an {{Online Business Writing Class}}}.
\newblock \bibinfo{journal}{\emph{American Journal of Distance Education}}
  \bibinfo{volume}{20}, \bibinfo{number}{2} (\bibinfo{date}{June}
  \bibinfo{year}{2006}), \bibinfo{pages}{93--107}.
\newblock
\showISSN{0892-3647, 1538-9286}
\urldef\tempurl%
\url{https://doi.org/10.1207/s15389286ajde2002_4}
\showDOI{\tempurl}


\bibitem[\protect\citeauthoryear{McDonald, Edwards, and Zhao}{McDonald
  et~al\mbox{.}}{2012}]%
        {mcdonald_exploring_2012}
\bibfield{author}{\bibinfo{person}{Sharon McDonald}, \bibinfo{person}{Helen~M.
  Edwards}, {and} \bibinfo{person}{Tingting Zhao}.}
  \bibinfo{year}{2012}\natexlab{}.
\newblock \showarticletitle{Exploring {{Think}}-{{Alouds}} in {{Usability
  Testing}}: {{An International Survey}}}.
\newblock \bibinfo{journal}{\emph{IEEE Transactions on Professional
  Communication}} \bibinfo{volume}{55}, \bibinfo{number}{1}
  (\bibinfo{date}{March} \bibinfo{year}{2012}), \bibinfo{pages}{2--19}.
\newblock
\showISSN{1558-1500}
\urldef\tempurl%
\url{https://doi.org/10.1109/TPC.2011.2182569}
\showDOI{\tempurl}


\bibitem[\protect\citeauthoryear{Mekler and Hornbæk}{Mekler and
  Hornbæk}{2016}]%
        {mekler_momentary_2016}
\bibfield{author}{\bibinfo{person}{Elisa~D. Mekler} {and}
  \bibinfo{person}{Kasper Hornbæk}.} \bibinfo{year}{2016}\natexlab{}.
\newblock \showarticletitle{Momentary Pleasure or Lasting Meaning?
  Distinguishing Eudaimonic and Hedonic User Experiences}. In
  \bibinfo{booktitle}{\emph{Proceedings of the 2016 {CHI} Conference on Human
  Factors in Computing Systems}} (New York, {NY}, {USA}).
  \bibinfo{publisher}{Association for Computing Machinery},
  \bibinfo{pages}{4509--4520}.
\newblock
\showISBNx{978-1-4503-3362-7}
\urldef\tempurl%
\url{https://doi.org/10.1145/2858036.2858225}
\showURL{%
\tempurl}


\bibitem[\protect\citeauthoryear{Microsoft}{Microsoft}{2021a}]%
        {microsoft_microsoft_2021-1}
\bibfield{author}{\bibinfo{person}{Microsoft}.}
  \bibinfo{year}{2021}\natexlab{a}.
\newblock \bibinfo{title}{Microsoft {{SharePoint}}: {{Build Team Intranets}} \&
  {{Share Files}}}.
\newblock
  \bibinfo{howpublished}{\url{https://www.microsoft.com/en-ca/microsoft-365/sharepoint/collaboration}}.
\newblock


\bibitem[\protect\citeauthoryear{Microsoft}{Microsoft}{2021b}]%
        {microsoft_microsoft_2021}
\bibfield{author}{\bibinfo{person}{Microsoft}.}
  \bibinfo{year}{2021}\natexlab{b}.
\newblock \bibinfo{title}{Microsoft {{Teams}}: {{Group Chat Software}}}.
\newblock
  \bibinfo{howpublished}{\url{https://www.microsoft.com/en-ca/microsoft-teams/group-chat-software}}.
\newblock


\bibitem[\protect\citeauthoryear{Minge and Th{\"u}ring}{Minge and
  Th{\"u}ring}{2018}]%
        {minge_mecue_2018}
\bibfield{author}{\bibinfo{person}{Michael Minge} {and}
  \bibinfo{person}{Manfred Th{\"u}ring}.} \bibinfo{year}{2018}\natexlab{}.
\newblock \showarticletitle{The {{MeCUE Questionnaire}} (2.0): {{Meeting Five
  Basic Requirements}} for {{Lean}} and {{Standardized UX Assessment}}}. In
  \bibinfo{booktitle}{\emph{Design, {{User Experience}}, and {{Usability}}:
  {{Theory}} and {{Practice}}}} \emph{(\bibinfo{series}{Lecture {{Notes}} in
  {{Computer Science}}})}, \bibfield{editor}{\bibinfo{person}{Aaron Marcus}
  {and} \bibinfo{person}{Wentao Wang}} (Eds.). \bibinfo{publisher}{{Springer
  International Publishing}}, \bibinfo{address}{{Cham}},
  \bibinfo{pages}{451--469}.
\newblock
\showISBNx{978-3-319-91797-9}
\urldef\tempurl%
\url{https://doi.org/10.1007/978-3-319-91797-9_33}
\showDOI{\tempurl}


\bibitem[\protect\citeauthoryear{Miro}{Miro}{2021}]%
        {miro_miro_2021}
\bibfield{author}{\bibinfo{person}{Miro}.} \bibinfo{year}{2021}\natexlab{}.
\newblock \bibinfo{title}{Miro: {{An Online Visual Collaboration Platform}} for
  {{Teamwork}}}.
\newblock \bibinfo{howpublished}{\url{https://miro.com/}}.
\newblock


\bibitem[\protect\citeauthoryear{MixPanel}{MixPanel}{2021}]%
        {mixpanel_2021}
\bibfield{author}{\bibinfo{person}{MixPanel}.} \bibinfo{year}{2021}\natexlab{}.
\newblock \bibinfo{title}{MixPanel: Product Analytics for Mobile, Web, \&
  More}.
\newblock \bibinfo{howpublished}{\url{https://mixpanel.com/}}.
\newblock


\bibitem[\protect\citeauthoryear{Molich}{Molich}{2011}]%
        {molich_quest_2011}
\bibfield{author}{\bibinfo{person}{Rolf Molich}.}
  \bibinfo{year}{2011}\natexlab{}.
\newblock \showarticletitle{The Quest for Quality: Usability Testing
  Assessment}. In \bibinfo{booktitle}{\emph{11th Danish Human-Computer
  Interaction Research Symposium}} (Frederiksberg, Denmark).
  \bibinfo{pages}{56--59}.
\newblock
\showISBNx{978-87-92524-17-1}


\bibitem[\protect\citeauthoryear{Molich, Dray, and Siegel}{Molich
  et~al\mbox{.}}{2004}]%
        {molich_tips_2004}
\bibfield{author}{\bibinfo{person}{Rolf Molich}, \bibinfo{person}{Susan Dray},
  {and} \bibinfo{person}{David Siegel}.} \bibinfo{year}{2004}\natexlab{}.
\newblock \showarticletitle{Tips and tricks for a better international
  usability test}. In \bibinfo{booktitle}{\emph{Extended abstracts of the 2004
  conference on Human factors and computing systems - {CHI} '04}} (Vienna,
  Austria). \bibinfo{publisher}{{ACM} Press}, \bibinfo{pages}{1606}.
\newblock
\showISBNx{978-1-58113-703-3}
\urldef\tempurl%
\url{https://doi.org/10.1145/985921.986168}
\showDOI{\tempurl}


\bibitem[\protect\citeauthoryear{Moran}{Moran}{2020}]%
        {moran_covid-19_2020}
\bibfield{author}{\bibinfo{person}{Kate Moran}.}
  \bibinfo{year}{2020}\natexlab{}.
\newblock \bibinfo{booktitle}{\emph{{COVID}-19 Has Changed Your Users}}.
\newblock
\urldef\tempurl%
\url{https://www.nngroup.com/articles/covid-changed-users/}
\showURL{%
\tempurl}


\bibitem[\protect\citeauthoryear{Moran and Pernice}{Moran and Pernice}{2020}]%
        {moran_remote_2020}
\bibfield{author}{\bibinfo{person}{Kate Moran} {and} \bibinfo{person}{Kara
  Pernice}.} \bibinfo{year}{2020}\natexlab{}.
\newblock \bibinfo{title}{Remote {{Moderated Usability Tests}}: {{Why}} to {{Do
  Them}}}.
\newblock
  \bibinfo{howpublished}{\url{https://www.nngroup.com/articles/moderated-remote-usability-test-why/}}.
\newblock


\bibitem[\protect\citeauthoryear{Murtza, Monroe, and Youmans}{Murtza
  et~al\mbox{.}}{2017}]%
        {murtza_heuristic_2017}
\bibfield{author}{\bibinfo{person}{Rabia Murtza}, \bibinfo{person}{Stephen
  Monroe}, {and} \bibinfo{person}{Robert~J. Youmans}.}
  \bibinfo{year}{2017}\natexlab{}.
\newblock \showarticletitle{Heuristic Evaluation for Virtual Reality Systems}.
\newblock  \bibinfo{volume}{61}, \bibinfo{number}{1} (\bibinfo{year}{2017}),
  \bibinfo{pages}{2067--2071}.
\newblock
\showISSN{2169-5067}
\urldef\tempurl%
\url{https://doi.org/10.1177/1541931213602000}
\showDOI{\tempurl}
\newblock
\shownote{Publisher: {SAGE} Publications Inc.}


\bibitem[\protect\citeauthoryear{Nielsen}{Nielsen}{1992}]%
        {nielsen_finding_1992}
\bibfield{author}{\bibinfo{person}{Jakob Nielsen}.}
  \bibinfo{year}{1992}\natexlab{}.
\newblock \showarticletitle{Finding Usability Problems through Heuristic
  Evaluation}. In \bibinfo{booktitle}{\emph{Proceedings of the {{SIGCHI
  Conference}} on {{Human Factors}} in {{Computing Systems}}}}
  \emph{(\bibinfo{series}{{{CHI}} '92})}. \bibinfo{publisher}{{Association for
  Computing Machinery}}, \bibinfo{address}{{New York, NY, USA}},
  \bibinfo{pages}{373--380}.
\newblock
\showISBNx{978-0-89791-513-7}
\urldef\tempurl%
\url{https://doi.org/10.1145/142750.142834}
\showDOI{\tempurl}


\bibitem[\protect\citeauthoryear{Nielsen}{Nielsen}{1994}]%
        {nielsen_10_1994}
\bibfield{author}{\bibinfo{person}{Jakob Nielsen}.}
  \bibinfo{year}{1994}\natexlab{}.
\newblock \bibinfo{title}{10 {{Usability Heuristics}} for {{User Interface
  Design}}}.
\newblock
  \bibinfo{howpublished}{\url{https://www.nngroup.com/articles/ten-usability-heuristics/}}.
\newblock


\bibitem[\protect\citeauthoryear{Nielsen and Landauer}{Nielsen and
  Landauer}{1993}]%
        {nielsen_mathematical_1993}
\bibfield{author}{\bibinfo{person}{Jakob Nielsen} {and}
  \bibinfo{person}{Thomas~K. Landauer}.} \bibinfo{year}{1993}\natexlab{}.
\newblock \showarticletitle{A Mathematical Model of the Finding of Usability
  Problems}. In \bibinfo{booktitle}{\emph{Proceedings of the {{INTERACT}} '93
  and {{CHI}} '93 {{Conference}} on {{Human Factors}} in {{Computing
  Systems}}}} \emph{(\bibinfo{series}{{{CHI}} '93})}.
  \bibinfo{publisher}{{Association for Computing Machinery}},
  \bibinfo{address}{{New York, NY, USA}}, \bibinfo{pages}{206--213}.
\newblock
\showISBNx{978-0-89791-575-5}
\urldef\tempurl%
\url{https://doi.org/10.1145/169059.169166}
\showDOI{\tempurl}


\bibitem[\protect\citeauthoryear{Nielsen and Mack}{Nielsen and Mack}{1994}]%
        {nielsen_usability_1994}
\bibfield{editor}{\bibinfo{person}{Jakob Nielsen} {and}
  \bibinfo{person}{Robert~L. Mack}} (Eds.). \bibinfo{year}{1994}\natexlab{}.
\newblock \bibinfo{booktitle}{\emph{Usability inspection methods}}.
\newblock \bibinfo{publisher}{John Wiley \& Sons, Inc.}
\newblock
\showISBNx{978-0-471-01877-3}


\bibitem[\protect\citeauthoryear{Noldus}{Noldus}{2020}]%
        {noldus_record_2020}
\bibfield{author}{\bibinfo{person}{Noldus}.} \bibinfo{year}{2020}\natexlab{}.
\newblock \bibinfo{title}{Record \& Annotate - {{Recording}} Options and Easy
  Annotation}.
\newblock
  \bibinfo{howpublished}{\url{https://www.noldus.com/viso/record-annotate}}.
\newblock


\bibitem[\protect\citeauthoryear{N{\o}rgaard and Hornb{\ae}k}{N{\o}rgaard and
  Hornb{\ae}k}{2006}]%
        {norgaard_what_2006}
\bibfield{author}{\bibinfo{person}{Mie N{\o}rgaard} {and}
  \bibinfo{person}{Kasper Hornb{\ae}k}.} \bibinfo{year}{2006}\natexlab{}.
\newblock \showarticletitle{What Do Usability Evaluators Do in Practice? An
  Explorative Study of Think-Aloud Testing}. In
  \bibinfo{booktitle}{\emph{Proceedings of the 6th Conference on {{Designing
  Interactive}} Systems}} \emph{(\bibinfo{series}{{{DIS}} '06})}.
  \bibinfo{publisher}{{Association for Computing Machinery}},
  \bibinfo{address}{{New York, NY, USA}}, \bibinfo{pages}{209--218}.
\newblock
\showISBNx{978-1-59593-367-6}
\urldef\tempurl%
\url{https://doi.org/10.1145/1142405.1142439}
\showDOI{\tempurl}


\bibitem[\protect\citeauthoryear{Norman}{Norman}{2002}]%
        {norman_design_2002}
\bibfield{author}{\bibinfo{person}{Donald~A. Norman}.}
  \bibinfo{year}{2002}\natexlab{}.
\newblock \bibinfo{booktitle}{\emph{The {{Design}} of {{Everyday Things}}}}.
\newblock \bibinfo{publisher}{{Basic Books, Inc.}}, \bibinfo{address}{{USA}}.
\newblock
\showISBNx{978-0-465-06710-7}


\bibitem[\protect\citeauthoryear{Oviedo and Fox~Tree}{Oviedo and
  Fox~Tree}{2021}]%
        {oviedo_meeting_2021}
\bibfield{author}{\bibinfo{person}{Vanessa~Y. Oviedo} {and}
  \bibinfo{person}{Jean~E. Fox~Tree}.} \bibinfo{year}{2021}\natexlab{}.
\newblock \showarticletitle{Meeting by Text or Video-Chat: {{Effects}} on
  Confidence and Performance}.
\newblock \bibinfo{journal}{\emph{Computers in Human Behavior Reports}}
  \bibinfo{volume}{3} (\bibinfo{date}{Jan.} \bibinfo{year}{2021}),
  \bibinfo{pages}{100054}.
\newblock
\showISSN{2451-9588}
\urldef\tempurl%
\url{https://doi.org/10.1016/j.chbr.2021.100054}
\showDOI{\tempurl}


\bibitem[\protect\citeauthoryear{Oztekin, Delen, Turkyilmaz, and Zaim}{Oztekin
  et~al\mbox{.}}{2013}]%
        {oztekin_machine_2013-2}
\bibfield{author}{\bibinfo{person}{Asil Oztekin}, \bibinfo{person}{Dursun
  Delen}, \bibinfo{person}{Ali Turkyilmaz}, {and} \bibinfo{person}{Selim
  Zaim}.} \bibinfo{year}{2013}\natexlab{}.
\newblock \showarticletitle{A Machine Learning-Based Usability Evaluation
  Method for {{eLearning}} Systems}.
\newblock \bibinfo{journal}{\emph{Decision Support Systems}}
  \bibinfo{volume}{56}, \bibinfo{number}{C} (\bibinfo{date}{Dec.}
  \bibinfo{year}{2013}), \bibinfo{pages}{63--73}.
\newblock
\showISSN{0167-9236}


\bibitem[\protect\citeauthoryear{Patern{\`o}, Schiavone, and Conti}{Patern{\`o}
  et~al\mbox{.}}{2017}]%
        {paterno_customizable_2017}
\bibfield{author}{\bibinfo{person}{Fabio Patern{\`o}},
  \bibinfo{person}{Antonio~Giovanni Schiavone}, {and} \bibinfo{person}{Antonio
  Conti}.} \bibinfo{year}{2017}\natexlab{}.
\newblock \showarticletitle{Customizable Automatic Detection of Bad Usability
  Smells in Mobile Accessed Web Applications}. In
  \bibinfo{booktitle}{\emph{Proceedings of the 19th {{International
  Conference}} on {{Human}}-{{Computer Interaction}} with {{Mobile Devices}}
  and {{Services}}}} \emph{(\bibinfo{series}{{{MobileHCI}} '17})}.
  \bibinfo{publisher}{{Association for Computing Machinery}},
  \bibinfo{address}{{New York, NY, USA}}, \bibinfo{pages}{1--11}.
\newblock
\showISBNx{978-1-4503-5075-4}
\urldef\tempurl%
\url{https://doi.org/10.1145/3098279.3098558}
\showDOI{\tempurl}


\bibitem[\protect\citeauthoryear{Paternoster, Giardino, Unterkalmsteiner,
  Gorschek, and Abrahamsson}{Paternoster et~al\mbox{.}}{2014}]%
        {paternoster_software_2014}
\bibfield{author}{\bibinfo{person}{Nicolò Paternoster},
  \bibinfo{person}{Carmine Giardino}, \bibinfo{person}{Michael
  Unterkalmsteiner}, \bibinfo{person}{Tony Gorschek}, {and}
  \bibinfo{person}{Pekka Abrahamsson}.} \bibinfo{year}{2014}\natexlab{}.
\newblock \showarticletitle{Software development in startup companies: A
  systematic mapping study}.
\newblock  \bibinfo{volume}{56}, \bibinfo{number}{10} (\bibinfo{year}{2014}),
  \bibinfo{pages}{1200--1218}.
\newblock
\showISSN{0950-5849}
\urldef\tempurl%
\url{https://doi.org/10.1016/j.infsof.2014.04.014}
\showDOI{\tempurl}


\bibitem[\protect\citeauthoryear{{PlaybookUX}}{{PlaybookUX}}{2021}]%
        {playbookux_playbookux_2021}
\bibfield{author}{\bibinfo{person}{{PlaybookUX}}.}
  \bibinfo{year}{2021}\natexlab{}.
\newblock \bibinfo{booktitle}{\emph{{PlaybookUX}: Scalable User Testing \&
  Interview Software}}.
\newblock
\urldef\tempurl%
\url{https://www.playbookux.com/}
\showURL{%
\tempurl}


\bibitem[\protect\citeauthoryear{Sarvghad and Tory}{Sarvghad and Tory}{2015}]%
        {sarvghad_exploiting_2015}
\bibfield{author}{\bibinfo{person}{Ali Sarvghad} {and} \bibinfo{person}{Melanie
  Tory}.} \bibinfo{year}{2015}\natexlab{}.
\newblock \showarticletitle{Exploiting Analysis History to Support
  Collaborative Data Analysis}. In \bibinfo{booktitle}{\emph{Proceedings of the
  41st {{Graphics Interface Conference}}}} \emph{(\bibinfo{series}{{{GI}}
  '15})}. \bibinfo{publisher}{{Canadian Information Processing Society}},
  \bibinfo{address}{{CAN}}, \bibinfo{pages}{123--130}.
\newblock
\showISBNx{978-0-9947868-0-7}


\bibitem[\protect\citeauthoryear{Sarvghad, Tory, and Mahyar}{Sarvghad
  et~al\mbox{.}}{2017}]%
        {sarvghad_visualizing_2017}
\bibfield{author}{\bibinfo{person}{Ali Sarvghad}, \bibinfo{person}{Melanie
  Tory}, {and} \bibinfo{person}{Narges Mahyar}.}
  \bibinfo{year}{2017}\natexlab{}.
\newblock \showarticletitle{Visualizing {{Dimension Coverage}} to {{Support
  Exploratory Analysis}}}.
\newblock \bibinfo{journal}{\emph{IEEE transactions on visualization and
  computer graphics}} \bibinfo{volume}{23}, \bibinfo{number}{1}
  (\bibinfo{date}{Jan.} \bibinfo{year}{2017}), \bibinfo{pages}{21--30}.
\newblock
\showISSN{1941-0506}
\urldef\tempurl%
\url{https://doi.org/10.1109/TVCG.2016.2598466}
\showDOI{\tempurl}


\bibitem[\protect\citeauthoryear{Satybaldiev, Hevesi, Hirsch, Rey, and
  Lukowicz}{Satybaldiev et~al\mbox{.}}{2019}]%
        {satybaldiev_coat_2019}
\bibfield{author}{\bibinfo{person}{Aziret Satybaldiev}, \bibinfo{person}{Peter
  Hevesi}, \bibinfo{person}{Marco Hirsch}, \bibinfo{person}{Vitor~Fortes Rey},
  {and} \bibinfo{person}{Paul Lukowicz}.} \bibinfo{year}{2019}\natexlab{}.
\newblock \showarticletitle{{{CoAT}}: A Web-Based, Collaborative Annotation
  Tool}. In \bibinfo{booktitle}{\emph{Adjunct {{Proceedings}} of the 2019 {{ACM
  International Joint Conference}} on {{Pervasive}} and {{Ubiquitous
  Computing}} and {{Proceedings}} of the 2019 {{ACM International Symposium}}
  on {{Wearable Computers}}}} \emph{(\bibinfo{series}{{{UbiComp}}/{{ISWC}} '19
  {{Adjunct}}})}. \bibinfo{publisher}{{Association for Computing Machinery}},
  \bibinfo{address}{{New York, NY, USA}}, \bibinfo{pages}{814--818}.
\newblock
\showISBNx{978-1-4503-6869-8}
\urldef\tempurl%
\url{https://doi.org/10.1145/3341162.3345592}
\showDOI{\tempurl}


\bibitem[\protect\citeauthoryear{Sauer, Jeffery, Land, and Yetton}{Sauer
  et~al\mbox{.}}{2000}]%
        {sauer_effectiveness_2000}
\bibfield{author}{\bibinfo{person}{Chris Sauer}, \bibinfo{person}{D.~Ross
  Jeffery}, \bibinfo{person}{Lesley Land}, {and} \bibinfo{person}{Philip
  Yetton}.} \bibinfo{year}{2000}\natexlab{}.
\newblock \showarticletitle{The {{Effectiveness}} of {{Software Development
  Technical Reviews}}: {{A Behaviorally Motivated Program}} of {{Research}}}.
\newblock \bibinfo{journal}{\emph{IEEE Transactions on Software Engineering}}
  \bibinfo{volume}{26}, \bibinfo{number}{1} (\bibinfo{date}{Jan.}
  \bibinfo{year}{2000}), \bibinfo{pages}{1--14}.
\newblock
\showISSN{0098-5589}
\urldef\tempurl%
\url{https://doi.org/10.1109/32.825763}
\showDOI{\tempurl}


\bibitem[\protect\citeauthoryear{Sears}{Sears}{1997}]%
        {sears_heuristic_1997}
\bibfield{author}{\bibinfo{person}{Andrew Sears}.}
  \bibinfo{year}{1997}\natexlab{}.
\newblock \showarticletitle{Heuristic {{Walkthroughs}}: {{Finding}} the
  {{Problems Without}} the {{Noise}}}.
\newblock \bibinfo{journal}{\emph{International Journal of Human-Computer
  Interaction}} \bibinfo{volume}{9}, \bibinfo{number}{3} (\bibinfo{date}{Sept.}
  \bibinfo{year}{1997}), \bibinfo{pages}{213--234}.
\newblock
\showISSN{1044-7318}
\urldef\tempurl%
\url{https://doi.org/10.1207/s15327590ijhc0903_2}
\showDOI{\tempurl}


\bibitem[\protect\citeauthoryear{Shi}{Shi}{2008}]%
        {shi_field_2008}
\bibfield{author}{\bibinfo{person}{Qingxin Shi}.}
  \bibinfo{year}{2008}\natexlab{}.
\newblock \showarticletitle{A Field Study of the Relationship and Communication
  between {{Chinese}} Evaluators and Users in Thinking Aloud Usability Tests}.
  In \bibinfo{booktitle}{\emph{Proceedings of the 5th {{Nordic}} Conference on
  {{Human}}-Computer Interaction Building Bridges - {{NordiCHI}} '08}}.
  \bibinfo{publisher}{{ACM Press}}, \bibinfo{address}{{Lund, Sweden}},
  \bibinfo{pages}{344}.
\newblock
\showISBNx{978-1-59593-704-9}
\urldef\tempurl%
\url{https://doi.org/10.1145/1463160.1463198}
\showDOI{\tempurl}


\bibitem[\protect\citeauthoryear{Silveira, Choma, Pereira, Guerra, and
  Zaina}{Silveira et~al\mbox{.}}{2021}]%
        {silveira_ux_2021}
\bibfield{author}{\bibinfo{person}{Sofia A.~M. Silveira},
  \bibinfo{person}{Joelma Choma}, \bibinfo{person}{Roberto Pereira},
  \bibinfo{person}{Eduardo~M. Guerra}, {and} \bibinfo{person}{Luciana A.~M.
  Zaina}.} \bibinfo{year}{2021}\natexlab{}.
\newblock \showarticletitle{{UX} Work in Software Start-Ups: Challenges from
  the Current State of Practice}. In \bibinfo{booktitle}{\emph{Agile Processes
  in Software Engineering and Extreme Programming}} (Cham)
  \emph{(\bibinfo{series}{Lecture Notes in Business Information Processing})},
  \bibfield{editor}{\bibinfo{person}{Peggy Gregory}, \bibinfo{person}{Casper
  Lassenius}, \bibinfo{person}{Xiaofeng Wang}, {and} \bibinfo{person}{Philippe
  Kruchten}} (Eds.). \bibinfo{publisher}{Springer International Publishing},
  \bibinfo{pages}{19--35}.
\newblock
\showISBNx{978-3-030-78098-2}
\urldef\tempurl%
\url{https://doi.org/10.1007/978-3-030-78098-2_2}
\showDOI{\tempurl}


\bibitem[\protect\citeauthoryear{Silverback}{Silverback}{2019}]%
        {silverback_silverback_2019}
\bibfield{author}{\bibinfo{person}{Silverback}.}
  \bibinfo{year}{2019}\natexlab{}.
\newblock \bibinfo{title}{Silverback 3}.
\newblock \bibinfo{howpublished}{\url{https://silverbackapp.com/}}.
\newblock


\bibitem[\protect\citeauthoryear{Simpson}{Simpson}{1991}]%
        {simpson_practice_1991}
\bibfield{author}{\bibinfo{person}{Mark Simpson}.}
  \bibinfo{year}{1991}\natexlab{}.
\newblock \showarticletitle{The Practice of Collaboration in Usability Test
  Design}.
\newblock  \bibinfo{volume}{38}, \bibinfo{number}{4} (\bibinfo{year}{1991}),
  \bibinfo{pages}{527--531}.
\newblock
\showISSN{0049-3155}
\urldef\tempurl%
\url{http://www.jstor.org/stable/43095829}
\showURL{%
\tempurl}
\newblock
\shownote{Publisher: Society for Technical Communication.}


\bibitem[\protect\citeauthoryear{Soure, Kuang, Fan, and Zhao}{Soure
  et~al\mbox{.}}{2021}]%
        {soure_coux_2021}
\bibfield{author}{\bibinfo{person}{Ehsan~Jahangirzadeh Soure},
  \bibinfo{person}{Emily Kuang}, \bibinfo{person}{Mingming Fan}, {and}
  \bibinfo{person}{Jian Zhao}.} \bibinfo{year}{2021}\natexlab{}.
\newblock \showarticletitle{{CoUX}: Collaborative Visual Analysis of
  Think-Aloud Usability Test Videos for Digital Interfaces}.
\newblock  (\bibinfo{year}{2021}), \bibinfo{pages}{1--11}.
\newblock
\showISSN{1941-0506}
\urldef\tempurl%
\url{https://doi.org/10.1109/TVCG.2021.3114822}
\showDOI{\tempurl}
\newblock
\shownote{Conference Name: {IEEE} Transactions on Visualization and Computer
  Graphics.}


\bibitem[\protect\citeauthoryear{Stern, Kelliher, Burleson, and
  Tolentino}{Stern et~al\mbox{.}}{2008}]%
        {stern_sharing_2008-1}
\bibfield{author}{\bibinfo{person}{Rebecca Stern}, \bibinfo{person}{Aisling
  Kelliher}, \bibinfo{person}{Winslow Burleson}, {and} \bibinfo{person}{Lisa
  Tolentino}.} \bibinfo{year}{2008}\natexlab{}.
\newblock \showarticletitle{Sharing the Squid: Tangible Workplace
  Collaboration}. In \bibinfo{booktitle}{\emph{{{CHI}} '08 {{Extended
  Abstracts}} on {{Human Factors}} in {{Computing Systems}}}}
  \emph{(\bibinfo{series}{{{CHI EA}} '08})}. \bibinfo{publisher}{{Association
  for Computing Machinery}}, \bibinfo{address}{{New York, NY, USA}},
  \bibinfo{pages}{3369--3374}.
\newblock
\showISBNx{978-1-60558-012-8}
\urldef\tempurl%
\url{https://doi.org/10.1145/1358628.1358859}
\showDOI{\tempurl}


\bibitem[\protect\citeauthoryear{Subramonyam, Drucker, and Adar}{Subramonyam
  et~al\mbox{.}}{2019}]%
        {subramonyam_affinity_2019}
\bibfield{author}{\bibinfo{person}{Hariharan Subramonyam},
  \bibinfo{person}{Steven Drucker}, {and} \bibinfo{person}{Eytan Adar}.}
  \bibinfo{year}{2019}\natexlab{}.
\newblock \showarticletitle{Affinity Lens: Data-Assisted Affinity Diagramming
  with Augmented Reality}.
\newblock
\urldef\tempurl%
\url{https://doi.org/10.1145/3290605.3300628}
\showDOI{\tempurl}


\bibitem[\protect\citeauthoryear{TechSmith}{TechSmith}{2020}]%
        {techsmith_morae_2020}
\bibfield{author}{\bibinfo{person}{TechSmith}.}
  \bibinfo{year}{2020}\natexlab{}.
\newblock \bibinfo{title}{Morae 3 {{Tutorials}}}.
\newblock
  \bibinfo{howpublished}{\url{https://www.techsmith.com/tutorial-morae-current.html}}.
\newblock


\bibitem[\protect\citeauthoryear{Toh}{Toh}{2021}]%
        {toh_colour_2021}
\bibfield{author}{\bibinfo{person}{Paul Toh}.} \bibinfo{year}{2021}\natexlab{}.
\newblock \bibinfo{title}{Colour schemes and templates}.
\newblock
  \bibinfo{howpublished}{\url{https://personal.sron.nl/~pault/\#sec:diverging}}.
\newblock


\bibitem[\protect\citeauthoryear{Tuch, Roth, {Hornbæk}, Opwis, and
  Bargas-Avila}{Tuch et~al\mbox{.}}{2012}]%
        {tuch_is_2012}
\bibfield{author}{\bibinfo{person}{Alexandre~N. Tuch},
  \bibinfo{person}{Sandra~P. Roth}, \bibinfo{person}{Kasper {Hornbæk}},
  \bibinfo{person}{Klaus Opwis}, {and} \bibinfo{person}{Javier~A.
  Bargas-Avila}.} \bibinfo{year}{2012}\natexlab{}.
\newblock \showarticletitle{Is beautiful really usable? Toward understanding
  the relation between usability, aesthetics, and affect in {HCI}}.
\newblock  \bibinfo{volume}{28}, \bibinfo{number}{5} (\bibinfo{year}{2012}),
  \bibinfo{pages}{1596--1607}.
\newblock
\showISSN{0747-5632}
\urldef\tempurl%
\url{https://doi.org/10.1016/j.chb.2012.03.024}
\showDOI{\tempurl}


\bibitem[\protect\citeauthoryear{Useberry}{Useberry}{2021}]%
        {useberry_useberry_2021}
\bibfield{author}{\bibinfo{person}{Useberry}.} \bibinfo{year}{2021}\natexlab{}.
\newblock \bibinfo{title}{Useberry: {{Codeless}} Prototype Analytics}.
\newblock \bibinfo{howpublished}{\url{https://www.useberry.com/}}.
\newblock


\bibitem[\protect\citeauthoryear{UserTesting}{UserTesting}{2020}]%
        {usertesting_usertesting_2020}
\bibfield{author}{\bibinfo{person}{UserTesting}.}
  \bibinfo{year}{2020}\natexlab{}.
\newblock \bibinfo{title}{{{UserTesting}}: {{How}} to Analyze and Share
  Results}.
\newblock
  \bibinfo{howpublished}{\url{https://help.usertesting.com/hc/en-us/articles/115003371891-How-to-analyze-and-share-results}}.
\newblock


\bibitem[\protect\citeauthoryear{UserTesting}{UserTesting}{2021}]%
        {usertesting_usertesting_2021}
\bibfield{author}{\bibinfo{person}{UserTesting}.}
  \bibinfo{year}{2021}\natexlab{}.
\newblock \bibinfo{title}{{{UserTesting}}: {{The Human Insight Platform}}}.
\newblock \bibinfo{howpublished}{\url{https://www.usertesting.com/}}.
\newblock


\bibitem[\protect\citeauthoryear{UserZoom}{UserZoom}{2021}]%
        {userzoom_userzoom_2021}
\bibfield{author}{\bibinfo{person}{UserZoom}.} \bibinfo{year}{2021}\natexlab{}.
\newblock \bibinfo{title}{{{UserZoom}}: {{Actionable UX}} Insights for Better
  Digital Experiences}.
\newblock \bibinfo{howpublished}{\url{https://www.userzoom.com/}}.
\newblock


\bibitem[\protect\citeauthoryear{UXQB}{UXQB}{2021}]%
        {uxqb_international_2021}
\bibfield{author}{\bibinfo{person}{UXQB}.} \bibinfo{year}{2021}\natexlab{}.
\newblock \bibinfo{title}{International {{Usability}} and {{UX Qualification
  Board}}}.
\newblock \bibinfo{howpublished}{\url{https://uxqb.org/en/}}.
\newblock


\bibitem[\protect\citeauthoryear{{UXTesting}}{{UXTesting}}{2022}]%
        {uxtesting_uxtesting_2022}
\bibfield{author}{\bibinfo{person}{{UXTesting}}.}
  \bibinfo{year}{2022}\natexlab{}.
\newblock \bibinfo{booktitle}{\emph{{UXTesting}: The Best and Most Intelligent
  User Experience Solution for Enterprises}}.
\newblock
\urldef\tempurl%
\url{https://www.uxtesting.io/uxtesting}
\showURL{%
\tempurl}


\bibitem[\protect\citeauthoryear{van Schaik and Ling}{van Schaik and
  Ling}{2008}]%
        {van_schaik_modelling_2008}
\bibfield{author}{\bibinfo{person}{Paul van Schaik} {and}
  \bibinfo{person}{Jonathan Ling}.} \bibinfo{year}{2008}\natexlab{}.
\newblock \showarticletitle{Modelling user experience with web sites:
  Usability, hedonic value, beauty and goodness}.
\newblock  \bibinfo{volume}{20}, \bibinfo{number}{3} (\bibinfo{year}{2008}),
  \bibinfo{pages}{419--432}.
\newblock
\showISSN{0953-5438}
\urldef\tempurl%
\url{https://doi.org/10.1016/j.intcom.2008.03.001}
\showDOI{\tempurl}


\bibitem[\protect\citeauthoryear{Vredenburg, Mao, Smith, and Carey}{Vredenburg
  et~al\mbox{.}}{2002}]%
        {vredenburg_survey_2002}
\bibfield{author}{\bibinfo{person}{Karel Vredenburg}, \bibinfo{person}{Ji-Ye
  Mao}, \bibinfo{person}{Paul~W. Smith}, {and} \bibinfo{person}{Tom Carey}.}
  \bibinfo{year}{2002}\natexlab{}.
\newblock \showarticletitle{A Survey of User-Centered Design Practice}. In
  \bibinfo{booktitle}{\emph{Proceedings of the {{SIGCHI Conference}} on {{Human
  Factors}} in {{Computing Systems}}}} \emph{(\bibinfo{series}{{{CHI}} '02})}.
  \bibinfo{publisher}{{Association for Computing Machinery}},
  \bibinfo{address}{{New York, NY, USA}}, \bibinfo{pages}{471--478}.
\newblock
\showISBNx{978-1-58113-453-7}
\urldef\tempurl%
\url{https://doi.org/10.1145/503376.503460}
\showDOI{\tempurl}


\bibitem[\protect\citeauthoryear{Walther}{Walther}{1996}]%
        {walther_computer-mediated_1996}
\bibfield{author}{\bibinfo{person}{Joseph~B. Walther}.}
  \bibinfo{year}{1996}\natexlab{}.
\newblock \showarticletitle{Computer-{{Mediated Communication}}:
  {{Impersonal}}, {{Interpersonal}}, and {{Hyperpersonal Interaction}}}.
\newblock \bibinfo{journal}{\emph{Communication Research}}
  \bibinfo{volume}{23}, \bibinfo{number}{1} (\bibinfo{date}{Feb.}
  \bibinfo{year}{1996}), \bibinfo{pages}{3--43}.
\newblock
\showISSN{0093-6502}
\urldef\tempurl%
\url{https://doi.org/10.1177/009365096023001001}
\showDOI{\tempurl}


\bibitem[\protect\citeauthoryear{Xiao and Kim}{Xiao and Kim}{2018}]%
        {xiao_study_2018}
\bibfield{author}{\bibinfo{person}{Xin-Ting Xiao} {and}
  \bibinfo{person}{Seung-In Kim}.} \bibinfo{year}{2018}\natexlab{}.
\newblock \showarticletitle{A Study on the User Experience of Smart Speaker in
  China - Focused on Tmall Genie and Mi {AI} Speaker -}.
\newblock  \bibinfo{volume}{16}, \bibinfo{number}{10} (\bibinfo{year}{2018}),
  \bibinfo{pages}{409--414}.
\newblock
\showISSN{2713-6434}
\urldef\tempurl%
\url{https://doi.org/10.14400/JDC.2018.16.10.409}
\showDOI{\tempurl}
\newblock
\shownote{Publisher: The Society of Digital Policy and Management.}


\bibitem[\protect\citeauthoryear{Yang, Scuito, Zimmerman, Forlizzi, and
  Steinfeld}{Yang et~al\mbox{.}}{2018}]%
        {yang_investigating_2018}
\bibfield{author}{\bibinfo{person}{Qian Yang}, \bibinfo{person}{Alex Scuito},
  \bibinfo{person}{John Zimmerman}, \bibinfo{person}{Jodi Forlizzi}, {and}
  \bibinfo{person}{Aaron Steinfeld}.} \bibinfo{year}{2018}\natexlab{}.
\newblock \showarticletitle{Investigating {{How Experienced UX Designers
  Effectively Work}} with {{Machine Learning}}}. In
  \bibinfo{booktitle}{\emph{Proceedings of the 2018 {{Designing Interactive
  Systems Conference}}}} \emph{(\bibinfo{series}{{{DIS}} '18})}.
  \bibinfo{publisher}{{Association for Computing Machinery}},
  \bibinfo{address}{{New York, NY, USA}}, \bibinfo{pages}{585--596}.
\newblock
\showISBNx{978-1-4503-5198-0}
\urldef\tempurl%
\url{https://doi.org/10.1145/3196709.3196730}
\showDOI{\tempurl}


\bibitem[\protect\citeauthoryear{Yu, Zhou, Wang, and Zhao}{Yu
  et~al\mbox{.}}{2019}]%
        {yu_evaluation_2019}
\bibfield{author}{\bibinfo{person}{Mengli Yu}, \bibinfo{person}{Ronggang Zhou},
  \bibinfo{person}{Huiwen Wang}, {and} \bibinfo{person}{Weihua Zhao}.}
  \bibinfo{year}{2019}\natexlab{}.
\newblock \showarticletitle{An evaluation for {VR} glasses system user
  experience: The influence factors of interactive operation and motion
  sickness}.
\newblock   \bibinfo{volume}{74} (\bibinfo{year}{2019}),
  \bibinfo{pages}{206--213}.
\newblock
\showISSN{0003-6870}
\urldef\tempurl%
\url{https://doi.org/10.1016/j.apergo.2018.08.012}
\showDOI{\tempurl}


\bibitem[\protect\citeauthoryear{Zhao, Glueck, Isenberg, Chevalier, and
  Khan}{Zhao et~al\mbox{.}}{2018}]%
        {zhao_supporting_2018}
\bibfield{author}{\bibinfo{person}{Jian Zhao}, \bibinfo{person}{Michael
  Glueck}, \bibinfo{person}{Petra Isenberg}, \bibinfo{person}{Fanny Chevalier},
  {and} \bibinfo{person}{Azam Khan}.} \bibinfo{year}{2018}\natexlab{}.
\newblock \showarticletitle{Supporting {{Handoff}} in {{Asynchronous
  Collaborative Sensemaking Using Knowledge}}-{{Transfer Graphs}}}.
\newblock \bibinfo{journal}{\emph{IEEE Transactions on Visualization and
  Computer Graphics}} \bibinfo{volume}{24}, \bibinfo{number}{1}
  (\bibinfo{date}{Jan.} \bibinfo{year}{2018}), \bibinfo{pages}{340--350}.
\newblock
\showISSN{1941-0506}
\urldef\tempurl%
\url{https://doi.org/10.1109/TVCG.2017.2745279}
\showDOI{\tempurl}


\bibitem[\protect\citeauthoryear{Zoom}{Zoom}{2021}]%
        {zoom_video_2021}
\bibfield{author}{\bibinfo{person}{Zoom}.} \bibinfo{year}{2021}\natexlab{}.
\newblock \bibinfo{title}{Video {{Conferencing}}, {{Web Conferencing}},
  {{Webinars}}, {{Screen Sharing}} - {{Zoom}}}.
\newblock \bibinfo{howpublished}{\url{https://zoom.us/}}.
\newblock


\end{thebibliography}










\end{document}